\documentclass{egpubl}
\usepackage{sca2026}

\SpecialIssuePaper         % uncomment for final version of Computer Graphics Forum, special issue
\CGFccby
\usepackage[T1]{fontenc}
\usepackage{dfadobe}

\usepackage{cite}  % comment out for biblatex with backend=biber
\BibtexOrBiblatex
\electronicVersion
\PrintedOrElectronic

\ifpdf \usepackage[pdftex]{graphicx} \pdfcompresslevel=9
\else \usepackage[dvips]{graphicx} \fi

\usepackage{egweblnk}
\usepackage[ruled,linesnumbered]{algorithm2e}
\DontPrintSemicolon
\usepackage{amsmath}
\usepackage{amssymb}
\usepackage{bm}
\usepackage{cleveref}
\usepackage{mathrsfs}

\title[Adaptive Fluid Cohomology on Surfaces]%
      {Adaptive Fluid Cohomology on Surfaces}

\author[Bastian Abt, David Stotko, Nils Wandel, Reinhard Klein]
{Bastian Abt\orcid{0009-0000-4076-5209}
\quad David Stotko\orcid{0009-0008-2270-5710}
\quad Nils Wandel\orcid{0000-0002-7787-3622}
\quad Reinhard Klein\orcid{0000-0002-5505-9347}
\\
University of Bonn}

\begin{document}

%uncomment for using teaser
\teaser{
\includegraphics[width=0.99\linewidth]{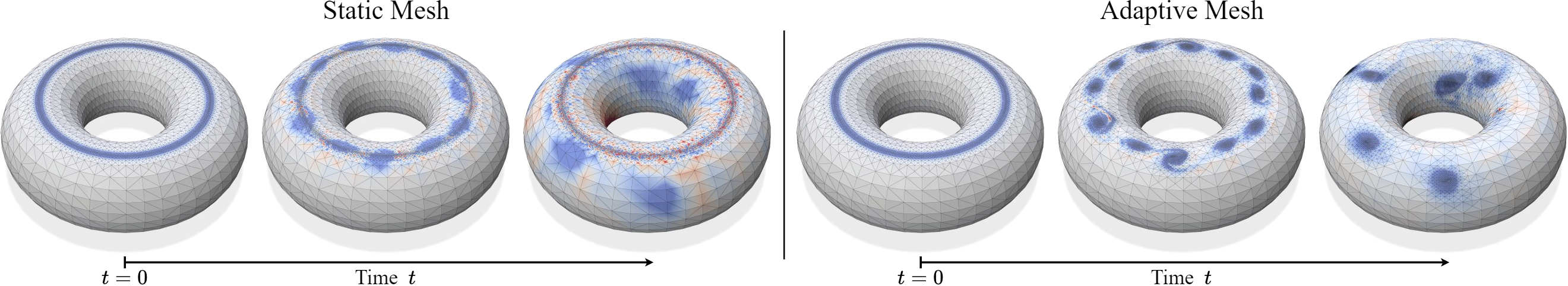}
\centering
\caption{Adaptive remeshing is crucial to ensure stable physical simulations with highly accurate results.
By incorporating temporal and spatial adaptivity into the original fluid cohomology framework (left), our method (right) resolves intricate and chaotic fluid flows while avoiding the memory footprint of globally refined meshes.
The image shows the vorticity of a simulation on a torus until the original/static method breaks due to numerical instabilities.}
\label{figure:teaser}
}

\maketitle
%-------------------------------------------------------------------------
\begin{abstract}
Simulating inviscid, incompressible fluids on non-simply-connected curved surfaces requires careful treatment of the flow's local and global behavior.
While recent theoretical advancements have established the critical dynamics of the harmonic component in such flows, practical applications remain computationally restricted by a lack of spatial and temporal adaptivity.
Furthermore, simulations on poor-quality meshes often lead to numerical instability and a failure to preserve the flow's underlying harmonic component when using naive interpolation methods.
In this paper, we introduce Adaptive Fluid Cohomology, a framework that integrates dynamic spatial and temporal refinement into the simulation of the Euler equations.
We leverage a posteriori error estimation to adjust spatial resolution on the fly, alongside a standard Dormand-Prince 5(4) time-stepping scheme for temporal accuracy.
To ensure stability during mesh mutations, we develop a novel method that robustly transfers the harmonic basis during remeshing.
While our experimental evaluation focuses on 2D surface flows, the underlying theoretical formulation is presented to capture the 3D setting as well.
Our evaluation demonstrates that this adaptive approach accurately recreates the dynamics of high-resolution simulations while reducing the memory footprint by up to $86\,\%$ and maintaining numerical stability even on poor-quality triangulations where static methods fail.

\begin{CCSXML}
<ccs2012>
    <concept>
        <concept_id>10010405.10010432</concept_id>
        <concept_desc>Applied computing~Physical sciences and engineering</concept_desc>
        <concept_significance>500</concept_significance>
    </concept>
    <concept>
        <concept_id>10010147.10010371.10010352.10010379</concept_id>
        <concept_desc>Computing methodologies~Physical simulation</concept_desc>
        <concept_significance>500</concept_significance>
    </concept>
    <concept>
        <concept_id>10002950.10003714</concept_id>
        <concept_desc>Mathematics of computing~Mathematical analysis</concept_desc>
        <concept_significance>300</concept_significance>
    </concept>
</ccs2012>
\end{CCSXML}

\ccsdesc[500]{Applied computing~Physical sciences and engineering}
\ccsdesc[500]{Computing methodologies~Physical simulation}
\ccsdesc[300]{Mathematics of computing~Mathematical analysis}

\printccsdesc
\end{abstract}
%-------------------------------------------------------------------------

\section{Introduction}

Computational fluid dynamics (CFD) remains a cornerstone of physics and computer science, enabling the realistic simulation of complex natural phenomena for engineering, weather analysis, and visual effects.
In the context of computer graphics, a particular emphasis is placed on simulating fluids governed by the inviscid, incompressible Euler equations on curved surfaces which are often represented as triangle meshes.
Traditional approaches often utilize the vorticity-streamfunction formulation, which is favored for its computational efficiency and its ability to naturally enforce the divergence-free constraint of idealized fluids.

While these methods are well-established for Euclidean domains, simulating flows on non-simply-connected curved manifolds introduces unique topological challenges.
Specifically, recovering the velocity field requires accounting for a harmonic component that represents the flow's global topology.
For years, this component was treated as static.
However, recent work by Yin~et~al.~\cite{yin_2023_fluid_cohomology} revealed that the harmonic part possesses its own critical dynamics.
This discovery provides the theoretical foundation for robust fluid solvers in non-simply-connected domains.

Despite these advancements, the practical application of fluid cohomology to complex, real-world animations is hindered by the lack of adaptivity.
The accuracy of these simulations is heavily dependent on mesh resolution and mesh quality in regions of high vorticity or rapid velocity changes.
In time-dependent simulations, these regions evolve, necessitating a framework that can dynamically refine and coarsen the discretization to maintain efficiency without sacrificing physical accuracy.
As we will show, naive interpolation of harmonic components during mesh modification often leads to numerical instability and a failure to preserve the underlying homology of the flow.

In this paper, we introduce Adaptive Fluid Cohomology, a framework that extends the work of Yin~et~al.~\cite{yin_2023_fluid_cohomology} by integrating both spatial and temporal adaptivity.
We employ the Dormand-Prince 5(4) Runge-Kutta method to ensure temporal accuracy.
For spatial adaptivity, we leverage intrinsic triangulations \cite{Sharp2019intrinsic,Gillespie2021intrinsic} and results of adaptive finite element methods (AFEM) to dynamically adjust the spatial resolution.
We introduce a novel topology-aware interpolation method for transferring harmonic forms during mesh adaptation.
By utilizing homology theory, we ensure that the discrete basis of harmonic forms reflects the same underlying continuous homology class even as the mesh is mutated.
Our evaluation demonstrates that this adaptive framework not only significantly improves numerical stability on irregular meshes but also preserves the physical behavior of the fluid more robustly than standard $L^2$-projection methods.
The full source code as well as our results are publicly available at \url{https://github.com/spookyGh0st/AdaptiveFluidCohomology.git}.

The main contributions of this work are summarized as follows:
\begin{itemize}
    \item We integrate spatial and temporal adaptivity into the fluid cohomology framework using a posteriori error estimation.
    \item We develop a robust method for transferring harmonic forms between different mesh resolutions that preserves the flow's homology class and compare it to the naive interpolation strategy.
    \item We provide an efficient implementation for 2D flows using intrinsic triangulation and demonstrate that
    \begin{itemize}
        \item our adaptive approach is able to recreate similar dynamics to a high-quality simulation while requiring comparable or reduced runtime as well as significantly reducing the memory footprint by up to $86\,\%$.
        \item our adaptive algorithm remains stable on poor-quality triangulations where the static method fails.
    \end{itemize}
\end{itemize}
\section{Related Work}

\subsection{Fluid Simulation Methods}

While the simulation of fluid motion in computer graphics is traditionally governed by the incompressible Navier-Stokes equations \cite{foster1997modeling,stam_1999_fluids}, researchers frequently employ and solve the Euler equations for inviscid phenomena such as smoke and gases.
Over the last decades, a large variety of simulation strategies were developed \cite{wang_2024_survey}.
Early development was based on grids by using the finite difference method \cite{harlow1962particle,foster1997modeling,forsythe1965finite,Yongning_2005_sand} and developed into the marker and cell method \cite{harlow1965numerical,Batty2007_solid_fluid}.
The Finite Volume Methods (FVM) \cite{LAI_1994_FVM,Moukalled2016,Dalal07082008} split the domain into discrete volumes of any grid type and approximate the surface and volume integrals of the governing equations.
Other methods like Smoothed Particle Hydrodynamics (SPH) \cite{Monaghan_2005,PRICE2012759,becker2007weakly,Solenthaler_2009_incompressible_sph,koschier2020smoothed,kelager2006lagrangian}, Fluid-Implicit-Particle (FLIP) \cite{BRACKBILL198825,Yongning_2005_sand,Ando_2011_sheets}, the Material Point Method (MPM) \cite{SULSKY1995236,Fei_2021_MPM,Stomakhin_2013_snow,Jiang_affine_PIC,Hu_2019_survey_MPM}, as well as Lattice Boltzmann Methods (LBM) \cite{Chen_1998_LBM,Aidun_2021_LBMcomplex,Li_2016_LBM,huang2015multiphase,guo2013lattice} shift the view to particle-based simulations and keep an underlying grid where it is advantageous.
Each method has distinct strengths depending on the application: while FLIP preserves fine turbulent detail with low numerical dissipation, MPM robustly handles solid–fluid coupling, and LBM handles porous media and intricate boundaries effectively.
Recently, machine learning techniques became increasingly popular by computing the whole resulting flow or by replacing parts in classical approaches \cite{Tompson_2017_accelerating,Dong_2019_nn,Sanchez_Gonzalez_2020_complex,wandel2021fluid}.

\subsection{Vorticity-Streamfunction Formulation}

The equations of fluid dynamics can be reformulated into the vorticity-streamfunction formulation in which the evolution of the vorticity is the main focus \cite{Fromm_1963_vorticity_streamfunction,Cheng1972numerical}.
The fluid velocity field is reconstructed from the streamfunction, serving as a vector potential, and inherently satisfies the usual incompressibility constraint \cite{Elcott2007simplicial}.
Significant advances have leveraged this property to capture highly detailed, turbulent dynamics without excessive grid resolution.
Foremost are vortex methods, which discretize the flow into Lagrangian primitives such as vortex particles \cite{selle2005vortex,meldgaard2022fast}, filaments \cite{weissmann2010filament,barnat2012smoke}, and sheets \cite{Pfaff2012sheets} to strictly minimize numerical dissipation.
Furthermore, other implementations have drastically improved computational efficiency and boundary handling through hybrid Eulerian-Lagrangian schemes \cite{Golas2012largescale,Park2005hybrid}, fast Poisson solvers, and impulse gauge methods \cite{Cortez1996impulse,feng2023impulse}, cementing vorticity-based frameworks as a staple for high-fidelity visual effects.
Despite their efficiency, these classical simulations do not capture the harmonic part of the flow in topologically non-trivial domains \cite{deGoes2016fields,Elcott2007simplicial}.
This limitation stems from the fact that a purely vorticity-driven reconstruction neglects the influence of non-trivial cohomology classes on the velocity field, particularly in the presence of obstacles or periodic boundaries \cite{Bhatia2013decomposition,yin_2023_fluid_cohomology}.
Following Yin~et~al.~\cite{yin_2023_fluid_cohomology}, traditional vorticity-streamfunction solvers and Biot-Savart methods are often topologically blind, either neglecting harmonic components entirely or inaccurately treating them as temporally static (e.g. \cite{Elcott2007simplicial}).
The fluid cohomology framework addresses this by deriving explicit equations of motion that allow these harmonic fields to dynamically evolve and interact with the local vorticity in complex domains.

\subsection{Adaptive Mesh Refinement in CFD}

Our adaptive approach builds upon a rich history of adaptive numerical methods in both computational mathematics and computer graphics.
Adaptive finite element methods (AFEM) and adaptive mesh refinement (AMR) have long been established as powerful tools for resolving complex multi-scale phenomena by minimizing linear system sizes and localizing computational effort \cite{bangerth2003adaptive,bonito2024adaptive}.
Foundational analyses of AFEM guarantee error reduction and optimal convergence through robust a posteriori error estimators \cite{carstensen2006convergence,verfurth2010posteriori,verfurth2013posteriori}.

These methods are essential for resolving multi-scale phenomena and steep flow gradients with optimal error bounds \cite{Blayo1999ocean,essadki2016adaptive,carstensen2006convergence}.
Furthermore, AMR is critical for tracking singularities in the vorticity-streamfunction formulation \cite{lal2024accuracy}.
To manage the performance demands of high-fidelity volumetric simulations, sophisticated dynamic data structures have been introduced, including unrestricted octrees \cite{Losasso2004octree}, adaptive tetrahedral meshes \cite{ando2013adaptive}, and sparse paged grids \cite{Setaluri2014adaptive,Raateland2022dcgrid}.

\section{Methodology}
\label{section:methodology}

This section presents the mathematical foundation of Eulerian fluid simulation on manifolds.
We first give a short summary of the fundamental equations of computational fluid simulation in \Cref{subsection:fluid_flows} based on Ferziger~et~al.~\cite{ferziger2019computational}.
Next, we translate these equations into the language of discrete exterior calculus in \Cref{subsection:covector_formulation}
and follow Yin~et~al.~\cite{yin_2023_fluid_cohomology} to develop the specifics of the fluid cohomology algorithm in \Cref{subsubsection:harmonic_forms}.
\Cref{subsubsection:discrete_harmonic_basis} describes the algorithm we employ to compute the necessary harmonic basis.
While our implementation operates in 2D only, we provide the general methodological foundation for analogous computations in higher dimensions.
Detailed descriptions regarding our contributions begin in \Cref{section:adaptive_simulation}.

\subsection{Fluid Flows}
\label{subsection:fluid_flows}

Fluids encompass liquids and gases and are characterized by the lack of resistance to deformation, which allows studying them as a continuous substance.
They are governed by conservation laws regarding mass and momentum.
These laws control the behavior of the following defining quantities, which are defined at each point in a spatial domain $M$.
The \textbf{velocity} field, denoted by $\bm{u}(x,t)$, is the velocity of the fluid at a fixed point $x$ in space at time $t$.
The mass \textbf{density} $\rho$ is the fluid's mass per unit volume.
Moreover, the \textbf{viscosity} measures the resistance of adjacent fluid layers to relative motion and enables further interaction of fluid particles similar to friction for solid objects.
Although we will only consider inviscid fluids, it should be noted that real fluids always have non-zero viscosity.

We first consider the conservation of mass which can be stated in its differential form as the continuity equation
\begin{equation}
    \frac{\partial}{\partial t} \rho + \mathrm{div}(\rho \bm{u}) = 0
\end{equation}
The conservation of momentum can be described using the stress tensor $T$ modeling the molecular rate of transport of momentum depending on the viscosity and $\bm{b}$ modeling the body forces per unit mass, e.g. gravity.
The resulting conservation law can be written as
\begin{equation}
    \frac{\partial}{\partial t} (\rho \bm{u}) + \mathrm{div}(\rho \bm{u} \otimes \bm{u}) = \mathrm{div}(T) + \rho \bm{b}
\end{equation}
where $\bm{u} \otimes \bm{u}$ is the outer product of the velocity with itself.

While the conservation equations are very general, we arrive at the Euler equations by using three approximations / assumptions \cite{ferziger2019computational}:
\begin{itemize}
    \item \textbf{Incompressibility:} the fluid density $\rho$ stays constant.
    This assumption is generally considered to be reasonable not only for liquids, but also for gases with low velocity.
    \item \textbf{Inviscid fluid:} the fluid viscosity is zero.
    Thus, the stress tensor reduces to $T = - p I$ governing only the volumetric stress based on the pressure $p$ and neglecting all distorting or shearing components.
    In simple terms, this means that we assume the fluid to have no internal "friction".
    \item \textbf{No body forces}: In many cases we neglect body forces, i.e. $\bm{b} = 0$, as influences like gravity have no significant effect in small-scale simulations.
\end{itemize}
If these assumptions hold true, the equations of conservation will reduce to
\begin{subequations}
\label{equation:euler_equations}
    \begin{alignat}{2}
        \frac{\partial}{\partial t} \bm{u} + (\bm{u} \cdot \nabla) \bm{u} & = - \frac{1}{\rho} \nabla p & \quad & \text{in } M \text{ (momentum equation)} \label{equation:euler_momentum} \\
        \mathrm{div}(\bm{u}) & = 0 & \quad & \text{in } M \text{ (divergence free)} \label{equation:euler_divergence_free} \\
        \langle \bm{u}, \bm{n} \rangle & = 0 & \quad & \text{on } \partial M \text{ (no-through boundary)}
    \end{alignat}
\end{subequations}
where $(\bm{u} \cdot \nabla) \bm{u}_i = \sum_j u_j \frac{\partial u_i}{\partial x_j}$ is the convective derivative and $\frac{p}{\rho}$ is often summarized as the kinematic pressure.
Because the fluid is incompressible, pressure and density are independent and only the gradient and not the absolute value of the pressure is physically meaningful.
These equations are independent of the dimension and therefore not only applicable for simulations in three dimensions but also on two-dimensional surfaces.

A particularly convenient mathematical formulation is obtained by introducing the \textbf{vorticity} $\bm{w}$ and the \textbf{streamfunction} $\bm{\psi}$.
The vorticity is defined as the curl of the velocity field yielding a scalar $w = \mathrm{curl}(\bm{u})$ for two-dimensional flows and a vector $\bm{w} = \mathrm{curl}(\bm{u})$ for three-dimensional flows.
The streamfunction is again a scalar field $\psi$ for 2D flows and a vector field $\bm{\psi}$ for 3D flows and serves as a vector potential for the velocity field:
\begin{subequations}
    \begin{align}
        u_x = \frac{\partial \psi}{\partial y} \enspace , \enspace u_y = - \frac{\partial \psi}{\partial x} & \qquad \text{in 2D} \\
        \bm{u} = \mathrm{curl}(\bm{\psi}) & \qquad \text{in 3D}
    \end{align}
\end{subequations}
Before explaining the benefits of using vorticity and streamfunction, we introduce the covector formulation of the Euler equations~\eqref{equation:euler_equations} \cite{Nabizadeh_2022_covector_fluids}.

\subsection{Covector Formulation}
\label{subsection:covector_formulation}

We briefly discuss the derivation of the covector form of \Cref{equation:euler_momentum} presented by Nabizadeh~et~al.~\cite{Nabizadeh_2022_covector_fluids}.
This allows us to study the fluid dynamics within the language of exterior calculus.
To do so, let $v := \bm{u}^\flat \in \Omega^1(M)$ denote the velocity 1-form and $p_\mathrm{L} = \frac{p}{\rho} - \frac{1}{2} |\bm{u}|^2 \in \Omega^0(M)$ the Lagrangian pressure.
After expanding \Cref{equation:euler_momentum} with $(\nabla \bm{u}) \cdot u = \frac{1}{2} \nabla |\bm{u}|^2$ and resulting in
\begin{equation}
    \frac{\partial}{\partial t} \bm{u} + (\bm{u} \cdot \nabla) \bm{u} + (\nabla \bm{u}) \cdot \bm{u} = - \frac{1}{\rho} \nabla p + \frac{1}{2} \nabla |\bm{u}|^2
\end{equation}
we can identify the Lie-derivative $\mathscr{L}_{\bm{u}}$ on the left-hand side and the Lagrangian pressure $p_\mathrm{L}$ on the right-hand side.
Additionally replacing the velocity field $\bm{u}$ with the velocity 1-form $v \in \Omega^1(M)$ and the gradient $\nabla$ with the exterior derivative $\mathrm{d}$ then yields the covector formulation:
\begin{subequations}
\label{equation:euler_covector_equations}
    \begin{alignat}{2}
        \frac{\partial}{\partial t} v + \mathscr{L}_{\bm{u}} v & = - \mathrm{d} p_\mathrm{L} & \quad & \text{in } M \text{ (circulation equation)} \label{equation:euler_covector_circulation} \\
        \delta v & = 0 & \quad & \text{in } M \text{ (co-closedness)} \label{equation:euler_covector_coclosedness} \\
        \iota^* \star v & = 0 & \quad & \text{on } \partial M \text{ (co-Dirichlet bound. cond.)} \label{equation:euler_co_boundary}
    \end{alignat}
\end{subequations}
The latter two equations express the divergence-free and no-through boundary conditions in covector form using the codifferential operator $\delta$, the pullback $^*$ of the canonical inclusion map $\iota : \partial M \hookrightarrow M$ and the Hodge star $\star$.

\begin{figure}[t]
    \centering
    \includegraphics[width=0.99\linewidth]{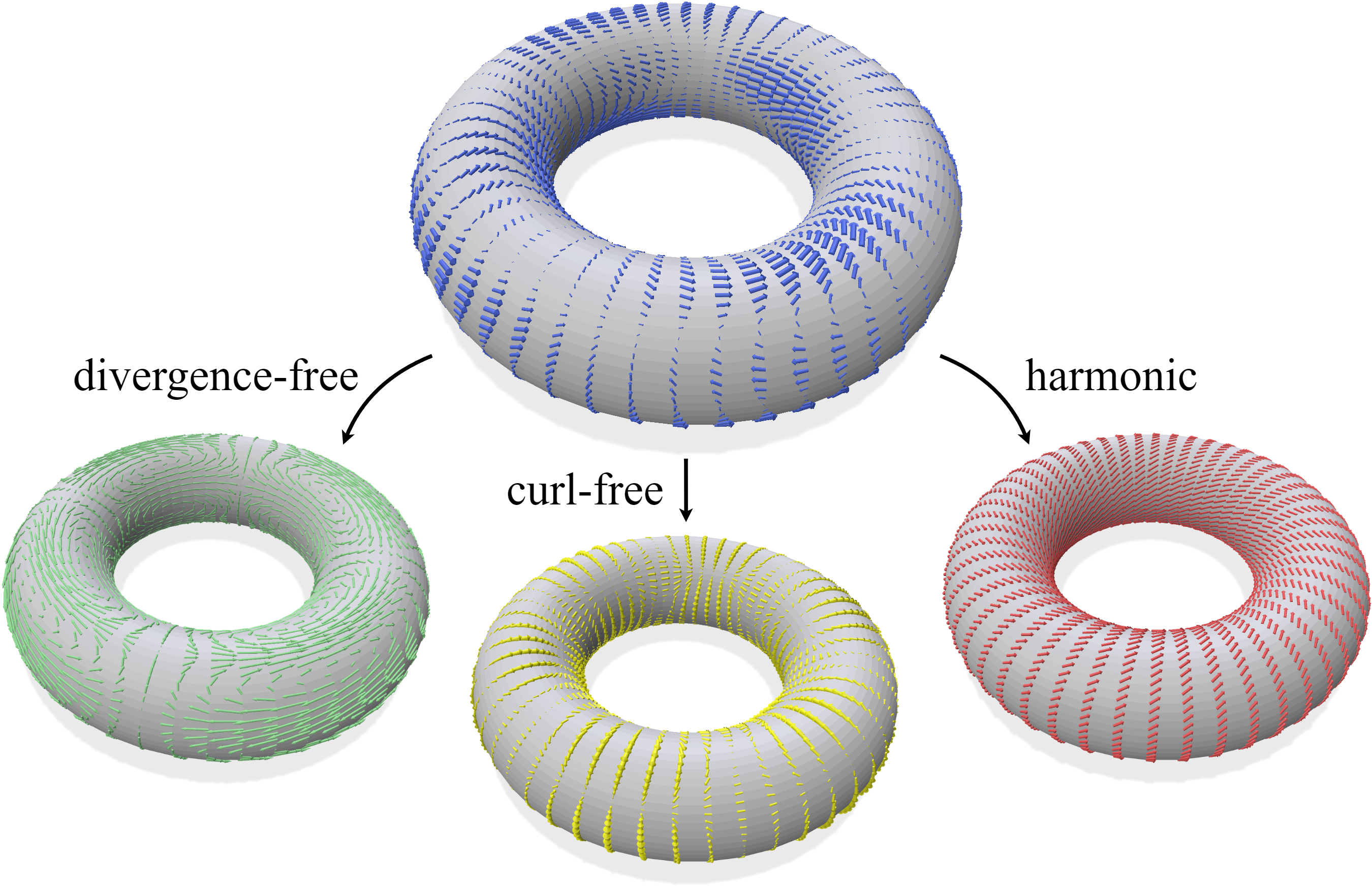}
    \caption{Helmholtz-Hodge decomposition of a vector field on a torus into divergence-free, curl-free and harmonic parts.}
    \label{figure:decomposition}
\end{figure}

The problem of solving Equations \eqref{equation:euler_covector_equations} becomes clearer when dividing the velocity 1-form into three orthogonal parts.
The Helmholtz-Hodge decomposition allows us to decompose any 1-form
\begin{equation}
\label{equation:hodge_decomposition}
    v = \mathrm{d} \alpha + \delta \beta + \gamma
\end{equation}
into the exterior derivative of a 0-form $\alpha \in \Omega^0(M)$, the codifferential of a 2-form $\beta \in \Omega^2(M)$ and a harmonic part $\gamma \in \mathcal{H}_\Delta^1(M)$ with $\mathrm{d} \gamma = 0, \delta \gamma = 0$ \cite{schwarz2006hodge}.
In vector calculus, we interpret these components as a divergence-free part $\mathrm{d} \alpha$, a curl-free part $\delta \beta$, and a harmonic part $\gamma$ as depicted in \Cref{figure:decomposition}.
The decomposition's orthogonality implies that $\delta \alpha \neq 0$ resulting in a violation of the co-closedness condition in \Cref{equation:euler_covector_coclosedness}.
As a consequence, we immediately obtain $\alpha = 0$ for any valid solution of the Euler equations.
Moreover, $\delta \beta$ and $\gamma$ fulfill the co-closedness condition as $\delta \delta \psi = 0$ and $\delta \gamma = 0$ by definition.
The remaining two terms $\beta$ and $\gamma$ are solved separately such that their combination recovers the full solution to the velocity field.

\subsubsection{Vorticity-Streamfunction Formulation}

Similar to the velocity, we can define a covector field for the vorticity as the 2-form $\omega = \mathrm{d} v$ \cite{yin_2023_fluid_cohomology}.
Applying the exterior derivative on \Cref{equation:euler_covector_circulation} and using the identity $\mathrm{d} \mathrm{d} = 0$ as well as the commutativity of the involved derivative operators, we obtain the vorticity equation
\begin{equation}
\label{equation:vorticity_equation}
    \frac{\partial}{\partial t} \omega + \mathscr{L}_{\bm{u}} \omega = 0
\end{equation}
To make \Cref{equation:vorticity_equation} more actionable, we can use the interior product $i_{\bm{u}}$ and Cartan's formula $\mathscr{L}_{\bm{u}} = i_{\bm{u}} \mathrm{d} + \mathrm{d} i_{\bm{u}}$ to replace the Lie derivative with the Lamb form $\lambda = - i_{\bm{u}} \omega$ \cite{yin_2023_fluid_cohomology}:
\begin{equation}
    \frac{\partial}{\partial t} \omega
    = - \mathscr{L}_{\bm{u}} \omega
    = - i_{\bm{u}} \underbrace{\mathrm{d} \omega}_{= \mathrm{d} \mathrm{d} v = 0} - \mathrm{d} i_{\bm{u}} \omega
    = - \mathrm{d} i_{\bm{u}} \omega
    = - \mathrm{d} \lambda
\end{equation}
In vector calculus, the corresponding lamb vector $\bm{l} = \lambda^\sharp$ can be calculated as $\bm{l} = \bm{w} \times \bm{u}$ with $\bm{w} = \mathrm{curl}(\bm{u})$ in 3D or the equivalent $l = (- u_y w, u_x w)$ in 2D.

We can recover one particular velocity 1-form by applying the Moore-Penrose pseudoinverse to the vorticity $v = \mathrm{d}^+ \omega$ defined by Yin~et~al.~\cite{yin_2023_fluid_cohomology}.
They also show that such a velocity 1-form is the codifferential $v = \delta \psi$ of a streamfunction 2-form $\psi$, \textbf{stream form} for short, that satisfies the co-Dirichlet boundary condition $\iota^* \star \psi = 0$ and takes the place of $\beta$ in the decomposition \eqref{equation:hodge_decomposition}.
Combining this with the definition of the vorticity leads to a Poisson problem for the stream form
\begin{equation}
\label{equation:stream_function_equation}
    \omega = \mathrm{d} v = \mathrm{d} \delta \psi
\end{equation}
Furthermore, Yin~et~al.~\cite{yin_2023_fluid_cohomology} prove that we can obtain a unique solution for the stream form by additionally requiring ${\iota^* \star \mathrm{d} \psi = 0}$ and $\psi \perp \mathcal{H}_\mathrm{C}^2(M)$, where $\mathcal{H}_\mathrm{C}^2(M)$ is the space of harmonic 2-forms that satisfy the co-Dirichlet boundary condition~\eqref{equation:euler_co_boundary}.
In vector calculus, the vorticity-streamfunction formulation guarantees the velocity field $\bm{u}$ to be divergence-free by construction without the need of additional projection steps.
The co-Dirichlet boundary condition is equivalent to requiring the streamfunction to vanish on the boundary.
This corresponds to $\psi = 0 \text{ on } \partial M$ in two dimensions, and $\bm{n} \times \bm{\psi} = 0 \text{ on } \partial M$ in three dimensions, where $\bm{n}$ denotes the boundary normal vector.

The main benefit of the vorticity-streamfunction formulation (besides ensuring the co-closedness) becomes apparent when we look at the special case of two-dimensional flows.
Here, vorticity and streamfunction are only one-dimensional quantities and the pressure term is eliminated in the equations which reduces the number of variables and differential equations to solve compared to the velocity-based formulation \cite{ferziger2019computational}.
We usually transform the vorticity and streamfunction into 0-forms $\hat{\omega} = \star \omega$ and $\hat{\psi} = \star \psi$ to make the calculations similar to the well-known approach using vector calculus.
For both the 2-forms and 0-forms, \Cref{equation:stream_function_equation} becomes the Poisson equation
\begin{equation}
\label{equation:poisson_equation}
    \omega = - \Delta \psi \qquad (\text{and similarly } \hat{\omega} = - \Delta \hat{\psi})
\end{equation}
with $\Delta = \frac{\partial^2}{\partial x^2} + \frac{\partial^2}{\partial y^2}$ being the Laplacian in 2D.

\subsubsection{Harmonic Forms}
\label{subsubsection:harmonic_forms}

The harmonic part $\gamma \in \mathcal{H}_\Delta^1(M)$ of the fluid velocity form has to fulfill the co-Dirichlet boundary condition \eqref{equation:euler_co_boundary} as well.
This reduces the complexity significantly, since the corresponding vector space~$\mathcal{H}_\mathrm{C}^1(M)$ is finite-dimensional \cite{schwarz2006hodge}.
As a consequence, we are able to choose a basis $\zeta^1, \dots, \zeta^n \in \mathcal{H}_\mathrm{C}^1(M)$ and a corresponding dual basis $\xi_1, \dots, \xi_n \in \mathcal{H}_\mathrm{C}^1(M)$ satisfying
\begin{equation}
    \int_M \zeta^i \wedge \xi_j = \delta^i_j
\end{equation}
with $\delta^i_j$ being the Kronecker delta.
We can describe the full velocity form of an arbitrary fluid flow as
\begin{equation}
    v = \delta \psi + \sum_{i = 1}^n c_i \zeta^i
\end{equation}
where the harmonic part is a linear combination with coefficients
\begin{equation}
    c_i = \int_M v \wedge \xi_i \in \mathbb{R}
\end{equation}
Thus, the dynamics now translate to a partial differential equation for the coefficients $c_i$, which was derived by Yin~et~al.~\cite{yin_2023_fluid_cohomology} to be
\begin{equation}
\label{equation:evolution_harmonic}
    \frac{\mathrm{d}}{\mathrm{d} t} c_i = \int_M \lambda \wedge \xi_i
\end{equation}
employing the lamb form $\lambda$.
When choosing an orthonormal basis $\zeta^1, \dots, \zeta^n \in \mathcal{H}_\mathrm{C}^1(M)$, the dual basis becomes $\xi_i = \star \zeta^i$ eliminating the need to track two different bases.
The necessary vector calculations also involve the vector counterpart $\bm{z}_i = (\zeta^i)^\sharp$ and read
\begin{equation}
    \frac{\mathrm{d}}{\mathrm{d} t} c_i = \int_M \bm{l} \cdot \bm{z}_i \, \mathrm{d} \bm{x}
\end{equation}
similarly for 2D and 3D.

\subsubsection{Discrete Harmonic Basis}
\label{subsubsection:discrete_harmonic_basis}

\begin{figure}[t]
    \centering
    \includegraphics[width=0.97\linewidth]{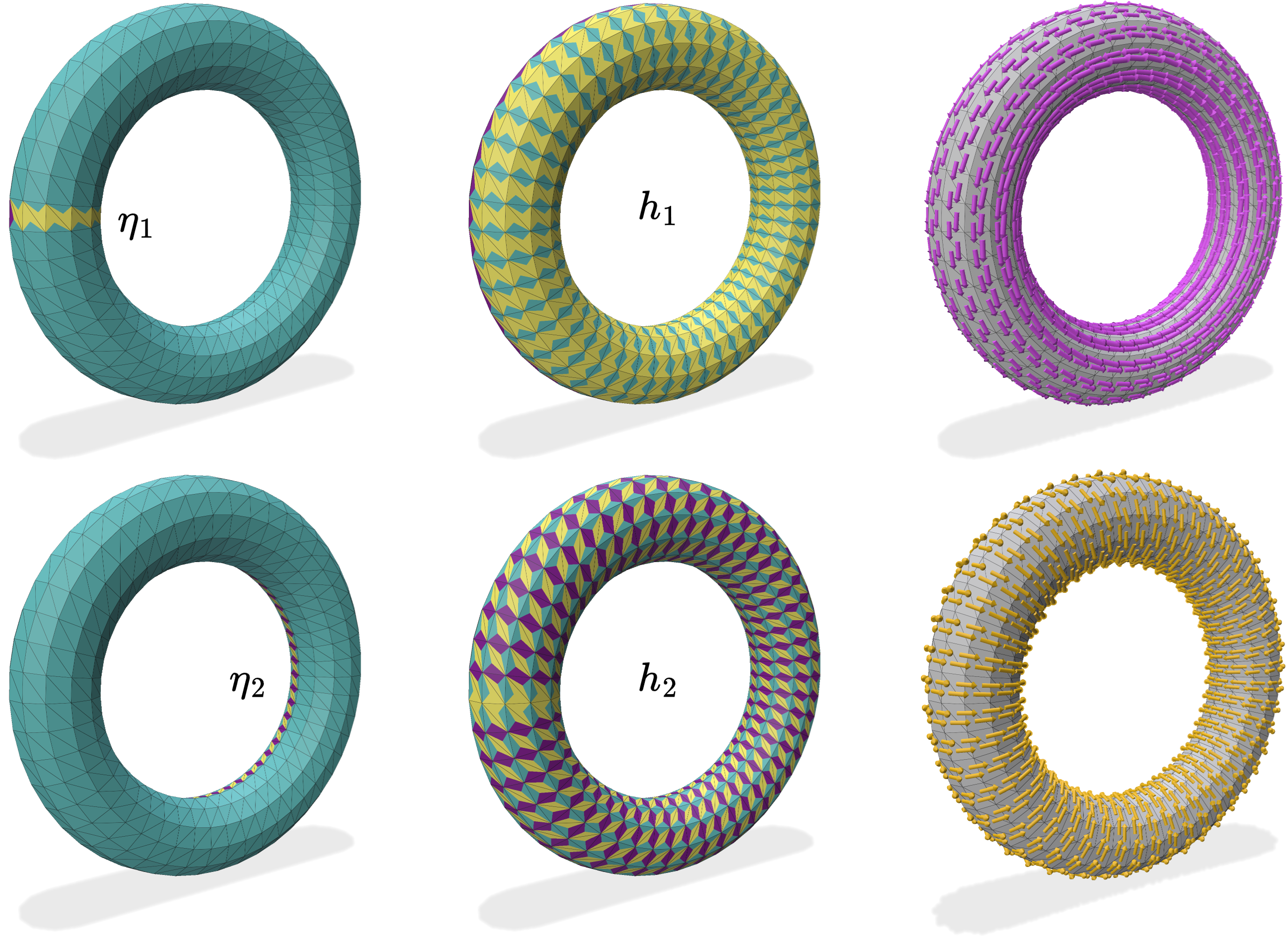}
    \caption{Construction of the two-dimensional harmonic basis of a torus.
    Initial paths and corresponding 1-forms $\eta_i$ (left) are projected onto harmonic forms $h_i$ (middle) and then converted into harmonic fields using Whitney Interpolation (right).}
    \label{figure:harmonic_field_creation}
\end{figure}

While we only describe the construction of the harmonic basis for 2-manifolds in detail within this section, an extension to higher dimensions follows naturally by increasing the dimension of the involved objects accordingly.
The practical construction of a manifold's harmonic basis relies on isolating the generators of the first (relative) homology group $H_1(M, \mathbb{Z})$.
For 2-manifolds, this is efficiently solved by applying a tree-cotree decomposition \cite{dlotko2012fast} to identify paths, the so-called (relative) \textbf{cycles}, within the manifold that serve as the generators for both the fundamental group (homotopy) and the first homology group $H_1(M, \mathbb{Z})$.
A path or cycle is described by selecting a set of connected edges in the dual mesh to form a loop or connect two boundaries (see left part of \Cref{figure:harmonic_field_creation}).
However, we can encode the path by saving the primal edges that are crossed by dual edges of the cycle.
While previously restricted to 2D surfaces \cite{erickson2005greedy,eppstein2002generators} and 3D volumes \cite{dey1998homology}, the generalization to arbitrary dimensions (involving a tri-partition of $m$-cells into an $m$-tree, an $m$-cotree, and a basis set) was developed in computational electromagnetics \cite{bossavit1998computational,Hiptmair_2002} and rigorously formalized for all polyhedral complexes by Edelsbrunner and Ölsböck \cite{edelsbrunner2020tri}.

Next, for each cycle representing a homology class, we compute its corresponding unique discrete harmonic 1-form, which represents the dual de Rham cohomology class \cite{crane2013digital}.
First, we build a discrete 1-form $\eta_i$ by setting each edge that is crossing the cycle $i$ to $+1$ or $-1$ depending on its crossing direction and all other edges to $0$, such that $\mathrm{d} \eta_i = 0$ (pseudocode is provided in \Cref{algorithm:delta_form}).
Next, we compute orthogonal projection of $\eta_i$ onto the kernel of $\delta$ by solving $(\mathrm{d}^T \mathrm{d}) \eta_{i,\mathrm{P}} = \mathrm{d}^T \eta_i$ and subtract this part from $\eta_i$.
The resulting form $h_i = \eta_i - \eta_{i,\mathrm{P}}$ is harmonic, as we first constructed a form with $\mathrm{d} \eta_i = 0$ and then subtracted the parts with $\delta \eta_i \neq 0$, only leaving the harmonic part (compare to the decomposition \eqref{equation:hodge_decomposition}).
The two initial 1-forms $\eta_1, \eta_2$ for a torus and their projections $h_1, h_2$ are depicted in \Cref{figure:harmonic_field_creation}.
Note that this harmonic form is the unique harmonic representative of each cohomology class.
We employ this uniqueness property in our adaptive algorithm to transfer the information between different mesh resolutions.

\subsubsection{Conversion to Vector Fields}

To conduct the necessary calculations for the lamb form and to represent the final flow field (or its components) on the manifold, we need to convert 1-forms to vector fields.
We employ Whitney-Interpolation for the conversion into vector fields and average them over each discrete volume to obtain a discretized vector field on the manifold.
The result of converting the harmonic basis of a torus into harmonic fields is shown in \Cref{figure:harmonic_field_creation}.

Finally, we also orthonormalize the harmonic basis using a $QR$ factorization with the custom inner product
\begin{equation}
    \langle a, b \rangle = \sum_{\text{Face } f} \langle a_f, b_f \rangle |f|
\end{equation}
with $|f|$ being the face's area.

\section{Adaptive Simulation}
\label{section:adaptive_simulation}

Building upon the geometric and topological foundations established in \Cref{section:methodology}, this section details our core technical contributions toward enabling fully dynamic spatiotemporal adaptivity for surface fluid simulations.
Specifically, \Cref{subsection:temporal_adaptivity} describes the temporal adaptivity framework, which employs a standard Dormand-Prince~5(4) time-stepping scheme \cite{DormandPrince1980}.
Alongside the established theoretical description, our spatial adaptivity utilizes a posteriori error estimation for the vorticity-streamfunction formulation (\Cref{subsection:spatial_adaptivity_vorticity}) and introduces our topology-aware transfer approach for the harmonic basis (\Cref{subsection:spatial_adaptivity_harmonic_basis}).
Finally, \Cref{subsection:newest_vertex_bisection} describes the newest vertex bisection strategy (NVB) \cite{mitchell1988unified,mitchell1989comparison} used for localized refinement and coarsening.
\Cref{section:implementation} further contains pseudocode for many of these algorithms.

\subsection{Temporal Adaptivity}
\label{subsection:temporal_adaptivity}

The equations of motion \eqref{equation:vorticity_equation} and \eqref{equation:evolution_harmonic} or their vector calculus counterparts allow us to apply standard discretization schemes to solve for the fluid flow.
It is well known that the stability of numerical solutions to partial differential equations depends on the ratio of spatial and temporal resolution and some simple problems even enable the calculation of precise mathematical bounds.
Thus, a spatially adaptive simulation directly motivates a temporally adaptive integration scheme to keep the simulation stable when high accuracy is needed and fast when the motion behaves well.
While Nabizadeh~et~al.~\cite{Nabizadeh_2022_covector_fluids} and Yin~et~al.~\cite{yin_2023_fluid_cohomology} employ Runge-Kutta~2 with back-and-forth error compensation and correction (BFECC) \cite{Dupont_BFECC,kim2005flowfixer} and Runge-Kutta~4, respectively, we choose the Dormand-Prince~5(4) method \cite{DormandPrince1980} to adapt the time step size $\Delta t$ automatically within a controllable window $[\Delta t_\mathrm{min}, \Delta t_\mathrm{max}]$.
Although the Dormand-Prince~5(4) method is a standard numerical choice in various simulation fields, it has not been utilized by the recent static frameworks we extend, where it is essential for unlocking temporal adaptivity.
Moreover, absolute and relative tolerances $\epsilon_\mathrm{abs}$ and $\epsilon_\mathrm{rel}$ as well as minimum and maximum step size scaling factors $f_\mathrm{min} \leq 1$ and $f_\mathrm{max} \geq 1$ allow the user to configure the temporal resolution easily with only a few parameters.
To control the step size, we estimate the local error as follows \cite{hairer1993solving}.
In addition to producing a numerical approximation $y$ of order $p$, a second approximation $\hat{y}$ of order $\hat{p}$ is constructed at every step.
As a measure of the error $e$, we calculate
\begin{equation}
    e = \sqrt{\frac{1}{n} \sum_{i = 1}^n \left( \frac{y_i - \hat{y}_i}{b_i} \right)^2}
\end{equation}
with $b_i = \epsilon_\mathrm{abs} + \epsilon_\mathrm{rel} \cdot \max(|y_i|, |\hat{y}_i|)$ being the controllable error bound.
Specifically for our fluid simulation, we combine all harmonic coefficients $c_i$ with $i \in \{1, \dots, n\}$ and the vorticity $\omega$ to compute
\begin{align*}
    y_i - \hat{y}_i & = c_i - \hat{c}_i \quad \forall i \in \{1, \dots, n\} \\
    y_{n+1} - \hat{y}_{n+1} & = \frac{1}{V_M} \int_M |\omega - \hat{\omega}|^2
\end{align*}
The new time step size $\Delta t'$ is then
\begin{equation*}
    \Delta t' = \Delta t \cdot \min \left( f_\mathrm{max}, \max \left( f_\mathrm{min}, \left( \frac{1}{e} \right)^{\frac{1}{q + 1}} \right) \right)
\end{equation*}
with $q = \min(p, \hat{p})$.
The time step size keeps getting updated iteratively until the step size is accepted by the condition $e \leq 1$.

\subsection{Spatial Adaptivity: Vorticity and Streamfunction}
\label{subsection:spatial_adaptivity_vorticity}

Our refinement strategies are based on results for the adaptive finite element method in which piecewise linear basis functions are usually employed.
We refer to the corresponding literature, e.g. \cite{schwarz2006hodge,verfurth2010posteriori,verfurth2013posteriori}, for the detailed mathematical statements and descriptions.

First, we consider the Poisson equation \eqref{equation:stream_function_equation} or \eqref{equation:poisson_equation} and the associated residual $R$ evaluated at an arbitrary auxiliary function $v$
\begin{equation}
    R(v) = \int_M \omega \, v - \int_M \nabla \psi \cdot \nabla v
\end{equation}
where $\omega$ and $\psi$ are the solutions to the Poisson problem in the AFEM sense.
The energy error between the FEM solution $\psi$ and the true continuous solution $\psi_\mathrm{true}$ satisfies
\begin{equation}
    \Vert \nabla (\psi_\mathrm{true} - \psi) \Vert = \underset{v \in H^1(M) \setminus \{0\}}{\mathrm{sup}} \frac{R(v)}{\Vert \nabla v \Vert}
\end{equation}
where $v$ is a function in Sobolev space $H^1(M)$.
The residual is then used to derive an upper bound~$\varepsilon_{R,V}$ on the error per volume element $V$ of the manifold.
An upper bound for the full manifold is obtained by summing over all discretization volumes $V$.
For surface meshes, we obtain the squared estimate
\begin{equation}
\label{equation:residual_error}
    \varepsilon_{R,f}^2 = h_f^2 \Vert \omega + \Delta \psi \Vert_f^2 + \frac{1}{2} \sum_{e \in \partial f} h_e \Vert \mathbb{J}_e(\bm{n}_e \cdot \nabla \psi) \Vert_e^2
\end{equation}
for each individual face $f$ with edges $e \in \partial f$.
Here, $h_f$ is the triangle diameter, $h_e$ the edge length, $\Vert \cdot \Vert_f$ and $\Vert \cdot \Vert_e$ the normalized integral (i.e. the "average value") of a quantity within a face $f$ or on edge $e$.
$\mathbb{J}_e(\cdot)$ denotes the jump value along $e$ originating from the piecewise linear description, making some quantities like $\nabla \psi$ discontinuous.

Given the residual $\varepsilon_{R,f}$ for each triangle, we can determine the subset of volume elements $V_\mathrm{R}$ to be refined and $V_\mathrm{C}$ to be coarsened.
We follow the Dörfler strategy \cite{doerfler_1996_convergent,stevenson2007optimality} where we define a threshold $\theta_\mathrm{R} \in (0, 1)$ to implicitly adjust the number of volumes to refine.
$V_\mathrm{R}$ is the set of minimal size that satisfies
\begin{equation}
\label{equation:doerfler_condition}
    \sum_{V \in V_\mathrm{R}} \varepsilon_{R,V}^2 \geq \theta_\mathrm{R} \sum_V \varepsilon_{R,V}^2
\end{equation}
We can compute $V_\mathrm{R}$ by sorting all volumes according to $\varepsilon_{R,V}^2$ and adding the ones with the largest error estimate to $V_\mathrm{R}$ until condition \eqref{equation:doerfler_condition} is satisfied.
Analogously, given a coarsening threshold $\theta_\mathrm{C} \in (0, 1)$, we select the set $V_\mathrm{C}$ of maximal size that satisfies $\sum_{V \in V_\mathrm{C}} \varepsilon_{R,V}^2 \leq \theta_\mathrm{C} \sum_V \varepsilon_{R,V}^2$ by adding elements from the sorted list, beginning with the smallest.
While positive properties like convergence and optimality are only proved for the refinement strategy, we choose this analogous strategy for the coarsening due to its synergistic definition and also expect it to inherit positive properties.
For practical reasons, we also include thresholds $\eta_\mathrm{R}$ and $\eta_\mathrm{C}$.
Triangles with residual $\varepsilon_{R,V} < \eta_\mathrm{R}$ will not be refined and triangles with $\varepsilon_{R,V} > \eta_\mathrm{C}$ will not be coarsened.
These parameters can be used to prevent the elements from refining and coarsening back and forth and give more control to the user in general.

This refinement and coarsening scheme is applied to the vorticity-streamfunction part of the fluid simulation.
We follow Gillespie~et~al.~\cite{Gillespie2021intrinsic} to transfer the vorticity values $\omega$ between different resolutions of the same mesh.
Starting with a base mesh $B$, we first refine the geometry and obtain a refined mesh $R$.
Afterwards, we apply the coarsening, which creates the final mesh $C$ of this remeshing step.
The $L^2$ distance between the solutions on the initial mesh $B$ and the final mesh $C$ is calculated as
\begin{equation}
\label{equation:transfer_equation}
    \Vert \omega_C - \omega_B \Vert_{L^2}^2 = (P_C \omega_C - P_B \omega_B)^T M_R (P_C \omega_C - P_B \omega_B)
\end{equation}
which uses $R$ as a common subdivision.
The computation requires mapping $\omega_B$ and $\omega_C$ linearly onto $R$ using interpolation matrices $P_B$ and $P_C$ and involves the Galerkin mass matrix $M_R$ of $R$.
The optimal solution of $\omega_C$ for minimizing \Cref{equation:transfer_equation} is given by the solution to the linear system
\begin{equation}
\label{equation:transfer_solution}
    (P_C^T M_R P_C) \omega_C = P_C^T M_R P_B \omega_B
\end{equation}
In the end, the streamfunction $\psi$ and the associated velocity field are recomputed from the vorticity on the remeshed geometry.

\subsection{Spatial Adaptivity: Harmonic Basis}
\label{subsection:spatial_adaptivity_harmonic_basis}

\begin{figure}[t]
    \centering
    \includegraphics[width=0.9\linewidth]{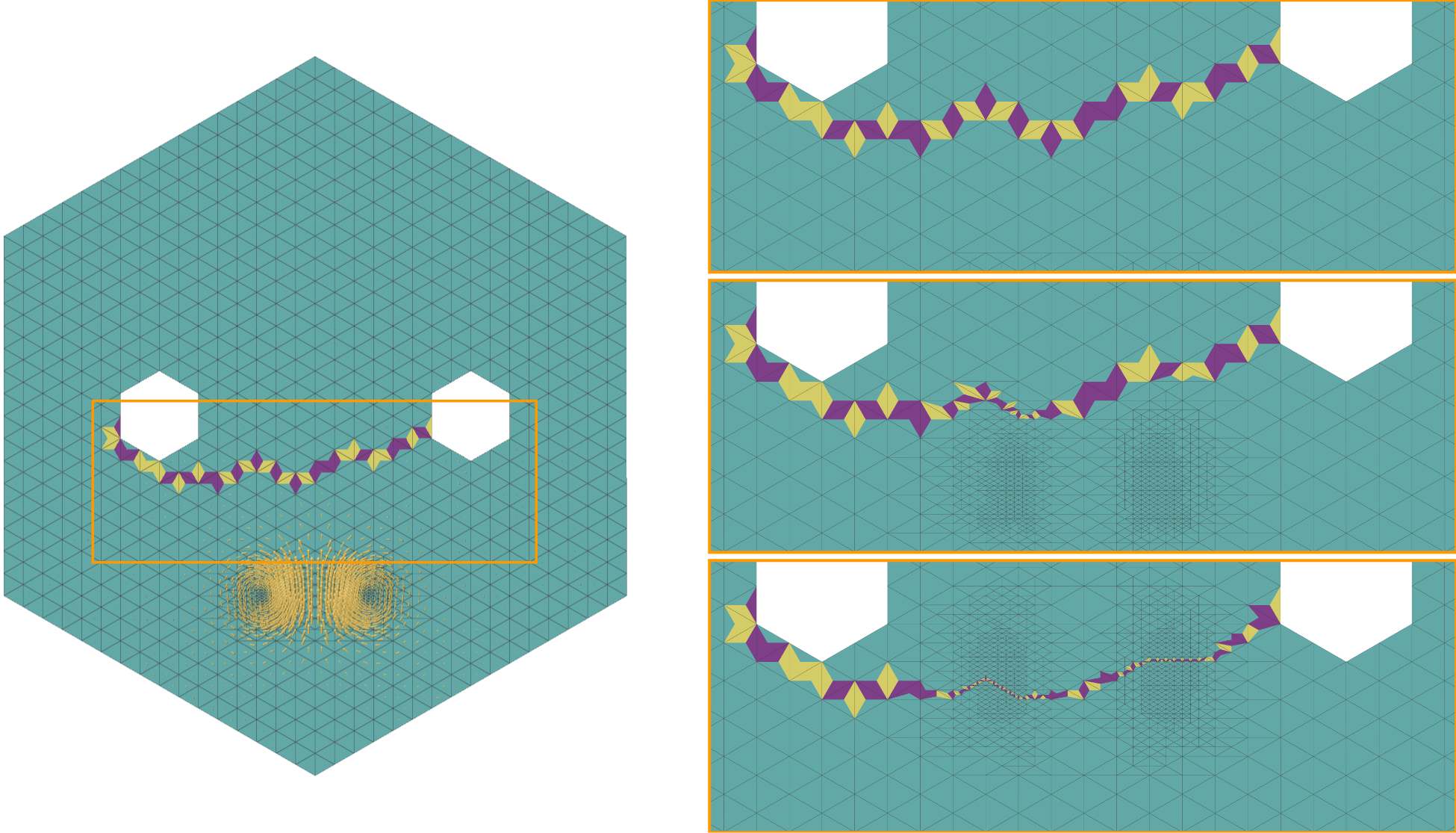}
    \caption{Path refinement for recomputing the harmonic basis.
    While two vortices (only shown left) pass through a path and cause refinement, the refined path stays within its original footprint (zoom in for details).
    The edge colors encode the property $\mathrm{d} \eta = 0$ (see \Cref{subsubsection:discrete_harmonic_basis}).}
    \label{figure:path_refinement}
\end{figure}

The harmonic part of the fluid flow consists of harmonic basis functions $\zeta^i$ and corresponding coefficients $c_i$.
The remeshing introduced in \Cref{subsection:spatial_adaptivity_vorticity} implies changes to the basis functions and in general to the coefficients as well.
However, we propose a method where the coefficients are simply reused and only a new consistent harmonic basis is computed.
We employ the theoretical results that directly connect cycles on the manifold (representing a homology class) to the unique harmonic representative of each cohomology class.
Hence, our goal is to change the existing cycles as little as possible.
Since we initially compute the cycles on the original unrefined mesh, we only need to handle the refinement of cycle faces because coarsening faces just reverts this process.
When a connected section of a cycle is refined, we keep track of all refined faces of the cycle and find a new path within the child faces to complete the cycle again.
We may choose an arbitrary path in case multiple possibilities exist.
We assign an arbitrary direction to the cycles in order to define whether a cycle enters or leaves a triangle through a given edge.
To select a specific cycle in the refined mesh, we employ a simple rule based on the half-edge data structure commonly used for surface meshes.
Whenever multiple path choices are available, we select the (half-)edge preceding the edge through which the cycle enters the current triangle.
The selected path may numerically influence the resulting harmonic basis, depending on the quality of the triangles traversed by the cycle, and a more sophisticated strategy could improve performance on poorly triangulated meshes.
Nevertheless, we did not observe any issues with the proposed scheme in our experiments.

This procedure, visualized in \Cref{figure:path_refinement}, creates new paths for the finer resolutions without changing the path at all from the coarser point of view.
Hence, we not only preserve the path that determines the harmonic basis but also reduce the computational costs by recomputing only the cycle segments impacted by remeshing.
The new cycle is then used to recalculate the harmonic basis function on the remeshed manifold.
During this step, we make use of the uniqueness property of the harmonic representative discussed in \Cref{subsubsection:discrete_harmonic_basis}.
While the near-original cycles guarantee the computation of the same harmonic representatives, the subsequent orthonormalization ensures consistent scaling (up to flipping signs, which is easy to handle) compared to the original harmonic basis.
In the end, we obtain discretizations of the same underlying continuous harmonic field for every triangulation created by the remeshing.
This strategy minimizes additional errors in the harmonic component by recomputing the harmonic basis directly on the updated mesh while preserving the underlying homology classes.
Since the harmonic coefficients are transferred without modification, the remeshing step itself does not introduce additional coefficient errors.
Consequently, the remaining inaccuracies are limited to the standard discretization errors inherent to the finite element representation on the adapted mesh.
In this sense, the proposed transfer procedure provides a principled foundation for minimizing harmonic reconstruction errors during adaptive remeshing while preserving circulation information consistently across mesh mutations.

Another approach is the transfer of harmonic fields using \Cref{equation:transfer_equation,equation:transfer_solution}.
For vector-valued functions like the harmonic basis fields, we solve these transfer equations component-wise.
However, in the special case of 2D manifolds with 2D vector fields, we interpret the tangent plane as the complex plane.
As a result, we only need to solve one complex-valued system of equations instead of two real-valued ones.
While this does not directly yield a harmonic field, restoring the harmonicity would result in the same basis as our previous approach while being computationally more expensive.
Thus, we will compare the accuracy of our recalculation approach against the interpolated approximation without doing the harmonic projection in \Cref{subsection:irregular_triangulation}.

\subsection{Newest Vertex Bisection (NVB)}
\label{subsection:newest_vertex_bisection}

\begin{figure}[t]
    \centering
    \includegraphics[width=0.99\linewidth]{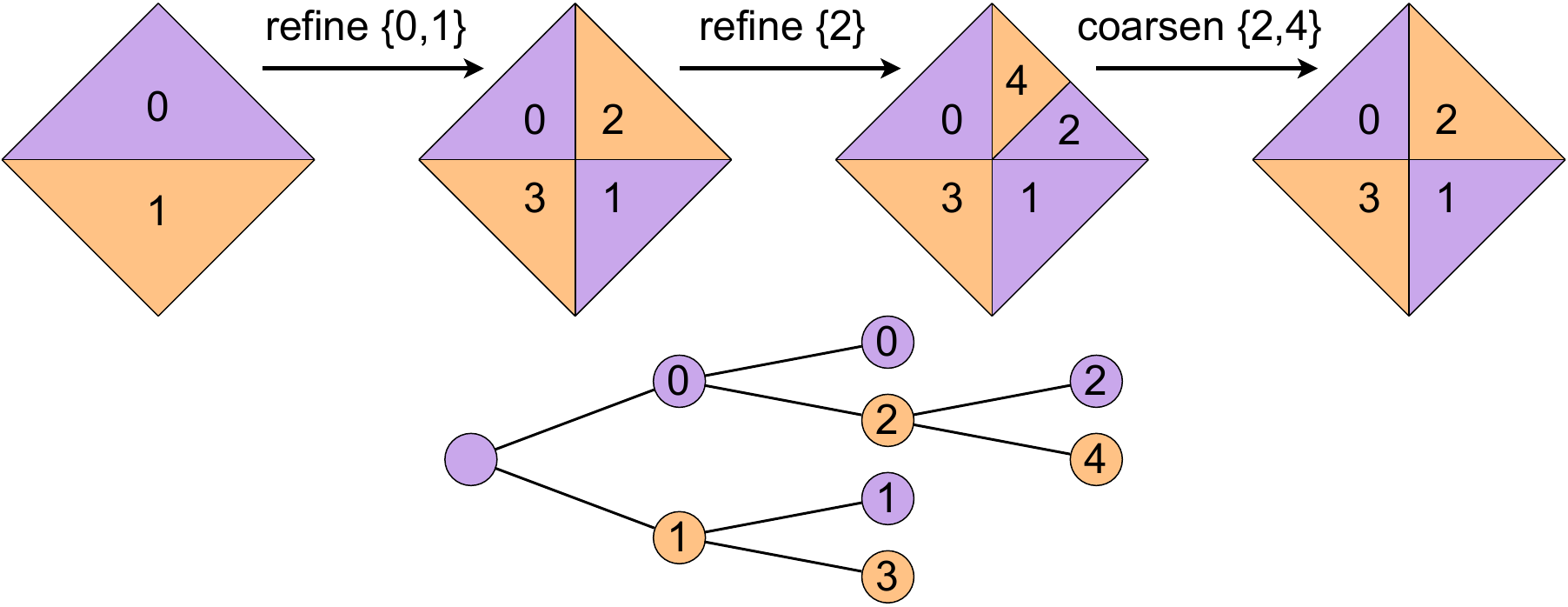}
    \caption{Face ordering in a sequence of refinement and coarsening steps.
        The implicit bisection tree is depicted below.}
    \label{figure:bisection_tree}
\end{figure}

We employ the newest vertex bisection strategy (NVB) \cite{mitchell1988unified,mitchell1989comparison} in our 2D implementation, whose general idea is as follows: Each face has a (initially defined) refinement edge.
This initialization is arbitrary but patterns resulting in uniform subdivisions or preserving the symmetry of a system may benefit the simulation.
Refining a face adds a vertex at the midpoint of the refinement edge and sets the refinement edge to the opposite edge of the newly created vertex.
To avoid hanging nodes, refinement is done iteratively until all faces are again triangles.

Classical implementations first refine marked faces and connect hanging nodes yielding a conforming triangulation.
However, in order to apply the NVB algorithm to an intrinsic triangulation \cite{Sharp2019intrinsic,Gillespie2021intrinsic}, we must ensure that the triangulation after each bisection is conforming.
In particular, we are not allowed to have temporary hanging nodes.
Thus, we focus on a different approach based on Chen and Zhang~\cite{chen2010coarsening}, which additionally gives rise to a coarsening algorithm as well.

We say an edge $e$ is \textbf{refinable} if $e$ is the refinement edge of all adjacent faces.
In this case we call the adjacent faces \textbf{compatible}.
Hence, we can insert a single new vertex in the edge center and refine both adjacent triangles.
Moreover, we call a vertex $v$ \textbf{coarsenable} if the refinement edges of all adjacent faces of $v$ lie opposite to~$v$.
This is the exact situation for a newly inserted vertex after a refinable edge is refined and therefore describes the inverse process.
In practice, we will not coarsen the geometry beyond its initial configuration.
The work of Chen and Zhang~\cite{chen2010coarsening} and the literature it is building on \cite{mitchell1988unified,Biedl2001efficient,KOSSACZKY1994recursive} provide theoretical grounding and practical solutions.
They prove that there exists a configuration in which all triangles are compatible as well as that refinements and coarsenings preserve compatibility.
In order to keep track of all triangles and their refinements, we employ a bisection tree that is implicitly stored by a face ordering \cite{chen2010coarsening}.
We illustrate the procedure to refine and coarsen triangles and the corresponding bisection tree in \Cref{figure:bisection_tree}.

\subsection{Adaptive Simulation}
\label{subsection:adaptive_simulation}

We discussed the individual parts of the simulation in detail within \Cref{section:methodology,section:adaptive_simulation}.
On the highest level (\Cref{algorithm:adaptive_fluid_solver}), the simulation consists of the initialization phase in which the cycles are built and the initial harmonic basis fields are computed.
Then, each simulation step consists of the spatial adaptation and the temporal integration.

The spatial adaptation itself is given in \Cref{algorithm:adapt_spatial_domain}.
This algorithm begins with the evaluation of the Dörfler strategy to perform the refinement and coarsening procedures afterwards.
We then transfer the vorticity to the new mesh, which concludes the adaptivity step for the vorticity component of the fluid flow.
In the end, we transfer the harmonic basis fields to the updated mesh by either recomputing or interpolating them.
As explained before, the harmonic coefficients are not changed during this step but only change according to the differential Equation~\ref{equation:evolution_harmonic} during (temporal) integration.

\begin{algorithm}[t]
    \caption{Adaptive Fluid Solver}
    \label{algorithm:adaptive_fluid_solver}
    \KwData{intrinsic triangulation $\mathcal{T}$, initial state $(\omega, c_1, \dots, c_m)$, initial $\Delta t \in \mathbb{R}$}
    $f_\mathrm{base} \gets$ suitable base face\;
    \tcp{$\bm{\sigma_i}$ and $c_i$ abbreviate all $m$ components}
    $\bm{\sigma}_i \gets$ \textsc{CyclesFromTreeCotree}($f_\mathrm{base}$) \tcp*{\cite{dlotko2012fast}}
    $\bm{h} \gets \textsc{HarmonicBasis}(\bm{\sigma}_i)$ \tcp*{\Cref{algorithm:harmonic_basis}}
    \ForEach{step}
    {
        $(\omega, c_i), \bm{h} \gets \textsc{AdaptSpatialDomain}(\mathcal{T}, \bm{\sigma}_i, \omega, c_i)$ \tcp*{\Cref{algorithm:adapt_spatial_domain}}
        $(\omega, c_i), \Delta t \gets \textsc{DormandPrince}((\omega, c_i), \bm{h}, \Delta t)$ \tcp*{\Cref{subsection:temporal_adaptivity}}
    }
\end{algorithm}

\begin{algorithm}[t]
    \caption{Adapt Spatial Domain}
    \label{algorithm:adapt_spatial_domain}
    \KwData{intrinsic triangulation $\mathcal{T}$, cycles $\bm{\sigma}_1, \dots, \bm{\sigma}_m$, initial state $(\omega, c_1, \dots, c_m)$}
    $\varepsilon_R \in \Omega^2 \gets$ Evaluate \Cref{equation:residual_error}\;
    $V_R, V_C \gets$ mark faces according to Dörfler strategy\;
    $\textsc{Refine}(\mathcal{T}, V_R)$, update $\bm{\sigma}_1, \dots, \bm{\sigma}_m$ \tcp*{\Cref{algorithm:refine}}
    $\textsc{Coarsen}(\mathcal{T}, V_C)$, update $\bm{\sigma}_1, \dots, \bm{\sigma}_m$ \tcp*{\Cref{algorithm:coarsen}}
    $\omega \gets$ transfer vorticity \tcp*{\Cref{subsection:spatial_adaptivity_vorticity}}
    $\bm{h} \gets \textsc{HarmonicBasis}(\bm{\sigma})$ \tcp*{\Cref{algorithm:harmonic_basis}}
    \Return{$(\omega, c_1, \dots, c_m), \bm{h}$}\;
\end{algorithm}

\section{Evaluation}

In this section we evaluate our adaptive method in a variety of test cases.
All results were run on a single thread on an AMD Ryzen 7 2700X Eight-Core Processor with a clock speed of 3700 MHz.
We build upon the geometry-central library \cite{geometrycentral}, which has been extended to provide all the necessary functionalities for our refinement strategy, and use the Eigen library for all linear algebra operations \cite{eigenweb}.
We provide videos of all evaluations as supplementary material.

In our evaluation examples, the initial conditions are set differently depending on the underlying mesh.
On planar meshes, we use one pair of Taylor-Green vortices by cropping the periodic functions accordingly.
For the (multi-)torus, the mesh is oriented horizontally and the flow is initialized according to vertical position: vertices on the upper portion of the torus are initialized with negative vorticity depicted in blue and rotating clockwise, while vertices on the lower portion are initialized with positive vorticity in blue and rotating counter-clockwise (e.g. see \Cref{figure:teaser}).
All parameters for the adaptivity calculations are given for each evaluation separately.
Recomputation of the harmonic basis is necessary after each remeshing process.
While it is possible to improve performance by remeshing only every $n$ simulation steps or adaptively based on velocity data or the time step size, we choose to remesh after every iteration (if necessary).
This shows that we are able to perform high-quality computations efficiently.
Additional experiments are provided in \Cref{section:extended_evaluation}.

\subsection{Regular Triangulation}

First, we compare the original method on a high resolution mesh of a good quality to our adaptive method using interpolated (nearly) harmonic fields and recomputed harmonic fields with our cycle refinement method.
We show the scalar vorticity field in regular intervals from 0 to 6 seconds as a clear visualization of the flow.

\begin{figure}[t]
    \centering
    \includegraphics[width=0.999\linewidth]{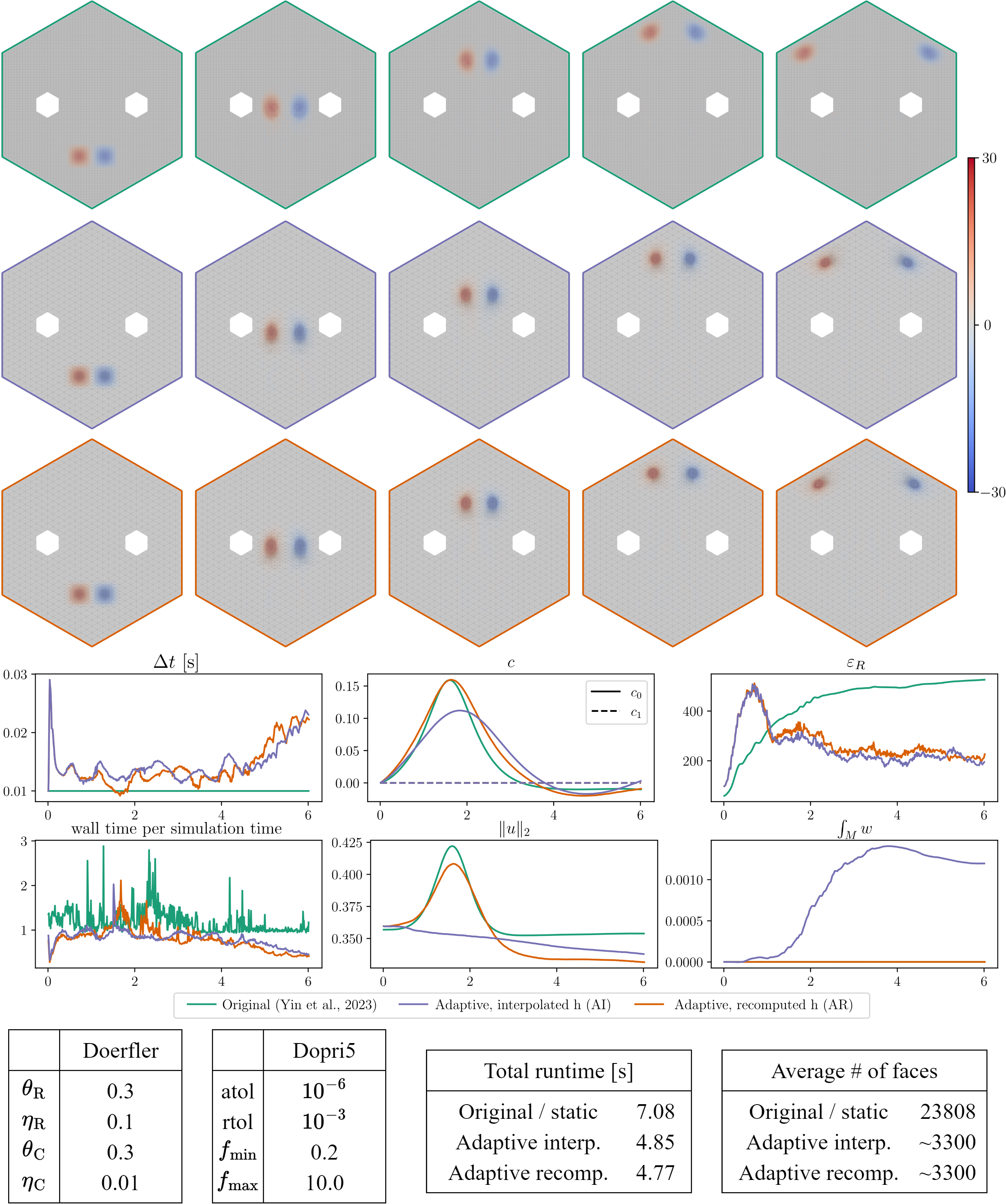}
    \caption{Simulation of a vortex pair on a hexagonal disc with two holes.
        We compare a high-quality static simulation (top, green border) with our adaptive solver using interpolated (middle, blue border) or recomputed (bottom, orange border) harmonic basis fields.}
    \label{figure:evaluation_basic}
\end{figure}

In \Cref{figure:evaluation_basic} we show how our adaptive solvers perform in comparison to the static method on a high resolution mesh of good quality with two holes.
In this case, the static method (top, green border) serves as a reference of a high-quality result.
In contrast, our adaptive methods only produce high spatial resolution in dynamic regions with large error residuals where also high vorticity is present.
This is sufficient to reproduce the overall motion very well.
Nonetheless, we can see that the vortices propagate slightly slower when simulated by the adaptive method with interpolated harmonic basis (middle, blue border), i.e. the vortices do not reach the same position after 6 seconds.
For our adaptive method with recomputed harmonic basis (bottom, orange border) this effect is less pronounced but still present.

We can also see in the two middle plots of \Cref{figure:evaluation_basic} that the harmonic coefficients $c$ and the average fluid velocity $\Vert u \Vert_2$ differ slightly.
Especially the interpolated basis prevents the flow from accelerating when the vortices are located between the holes causing a large qualitative difference.
The diagrams on the right depict the total residual estimate $\epsilon_R$ and the integrated vorticity $\int_M \omega$.
In this symmetric case the vorticity is supposed to contain equal amounts of positive and negative values and thus average to zero.
We can see that both adaptive methods first start with a large residual estimate $\epsilon_R$ but proceed to decrease it as the simulation continues.
In contrast, the residual estimate of the static method continuously increases and quickly exceeds the residual of both adaptive methods.
When we observe the integrated vorticity, only the version with interpolated harmonic basis deviates from zero and therefore from a perfectly symmetric flow.
The two left plots show the time step size $\Delta t$ and the wall time, which we normalize with the time step size.
Thus, wall times less than or equal to one indicate real-time performance.
The adaptive time stepping naturally decreases the time step size when the vortices are close to the holes and increases it afterwards, as the vortices propagate forward freely.
This has direct implications for the wall time.
Although spatial and temporal adaptivity require a substantial amount of additional computations, the reduced number of elements greatly improves the computation time of heavy computations like solving the Poisson problem \eqref{equation:poisson_equation}.
In total, the adaptive methods finish the six-second simulation in $4.8$ seconds while the static method needs $7.08$ seconds; a speed-up of $32\,\%$.
This efficiency is even more pronounced in terms of memory usage.
Instead of $23808$ triangles for the high-resolution static method, the adaptive ones are able to reduce this by $86\,\%$ to around $3300$ triangles with the given configuration.
Such a reduction also directly translates to the number of vertices, edges, all associated quantities and thus the sparse linear systems of equations.

\subsection{Irregular Triangulation}
\label{subsection:irregular_triangulation}

\begin{figure}[t]
    \centering
    \includegraphics[width=0.999\linewidth]{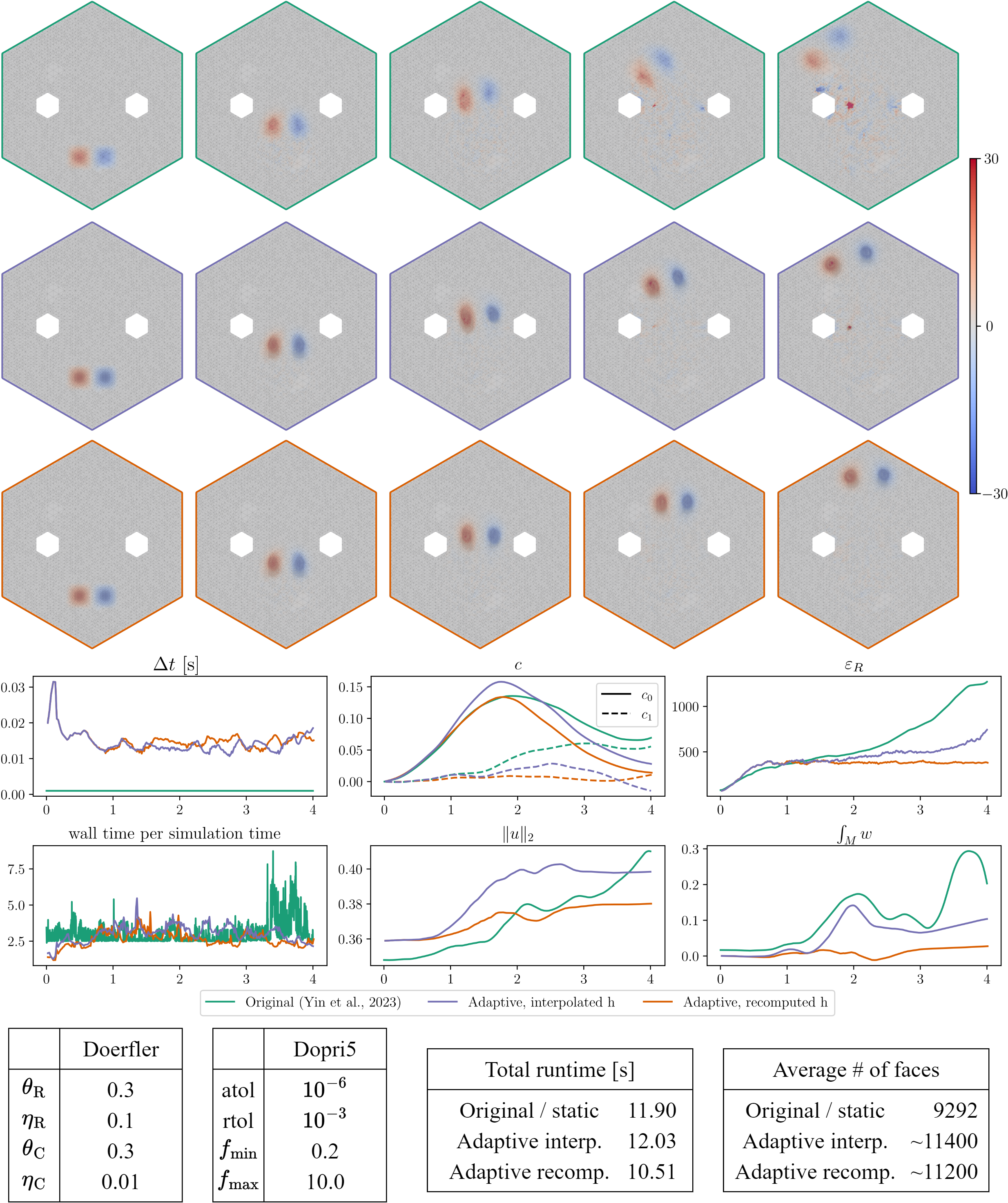}
    \caption{The same simulation of a vortex pair on an irregular mesh.
        The asymmetric mesh has a great influence on the static simulation (top, green border) and introduces an asymmetry as well as numerical instabilities in the simulation result.
        The interpolating adaptive method suffers from the same problems, although less pronounced.
        Only the adaptive method with recomputed basis shows almost no asymmetry nor instabilities.}
    \label{figure:evaluation_irregular_mesh}
\end{figure}

Despite the ability to control the resolution, adaptive methods start to shine in situations in which poor-quality meshes are initially given.
We construct such a mesh by assigning the initial refinement edge for the NVB randomly for each triangle.
Uniform refinements of all faces then create the mesh in \Cref{figure:evaluation_irregular_mesh} with $9292$ faces.
The static simulation suffers from the poor mesh quality such that the vortices are deflected to the left after passing through the two holes.
Moreover, numerical instabilities start to form and grow quickly until the whole simulation completely breaks despite using a really small time step size of $0.001$ seconds.
Therefore, \Cref{figure:evaluation_irregular_mesh} only shows the first four simulated seconds.
In contrast, both adaptive methods are able to perform stable simulations with lower error estimates for both the residual error $\varepsilon_R$ and the integrated vorticity.
Although the adaptive remeshing increases the number of faces compared to the static method (because we use the $9292$ faces as a base mesh that is not allowed to be coarsened), it does so by only $23\,\%$ or $21\,\%$ depending on the adaptive method.
Nonetheless, the discretization error of the interpolated harmonic basis also causes a left drift and instabilities are still visible.
The drifting in both simulations can be explained by a significant positive integrated vorticity $\int_M \omega$ creating net counter-clockwise rotation.
Both problems are eliminated by employing our recomputed harmonic basis, i.e. the vortices move straight forward and also no unstable regions formed.
These improvements are also visible in error metrics with the lowest residual $\varepsilon_R$ and almost zero integrated vorticity at all times.
Positive effects are also present as a $12\,\%$ reduction in total runtime and slightly fewer triangles compared to the other adaptive method.
This comparison not only demonstrates the advantages of adaptive methods in scenarios in which the static method fails but also highlights the importance of the harmonic components that we recompute reliably.

\begin{figure}[t]
    \centering
    \includegraphics[width=0.999\linewidth]{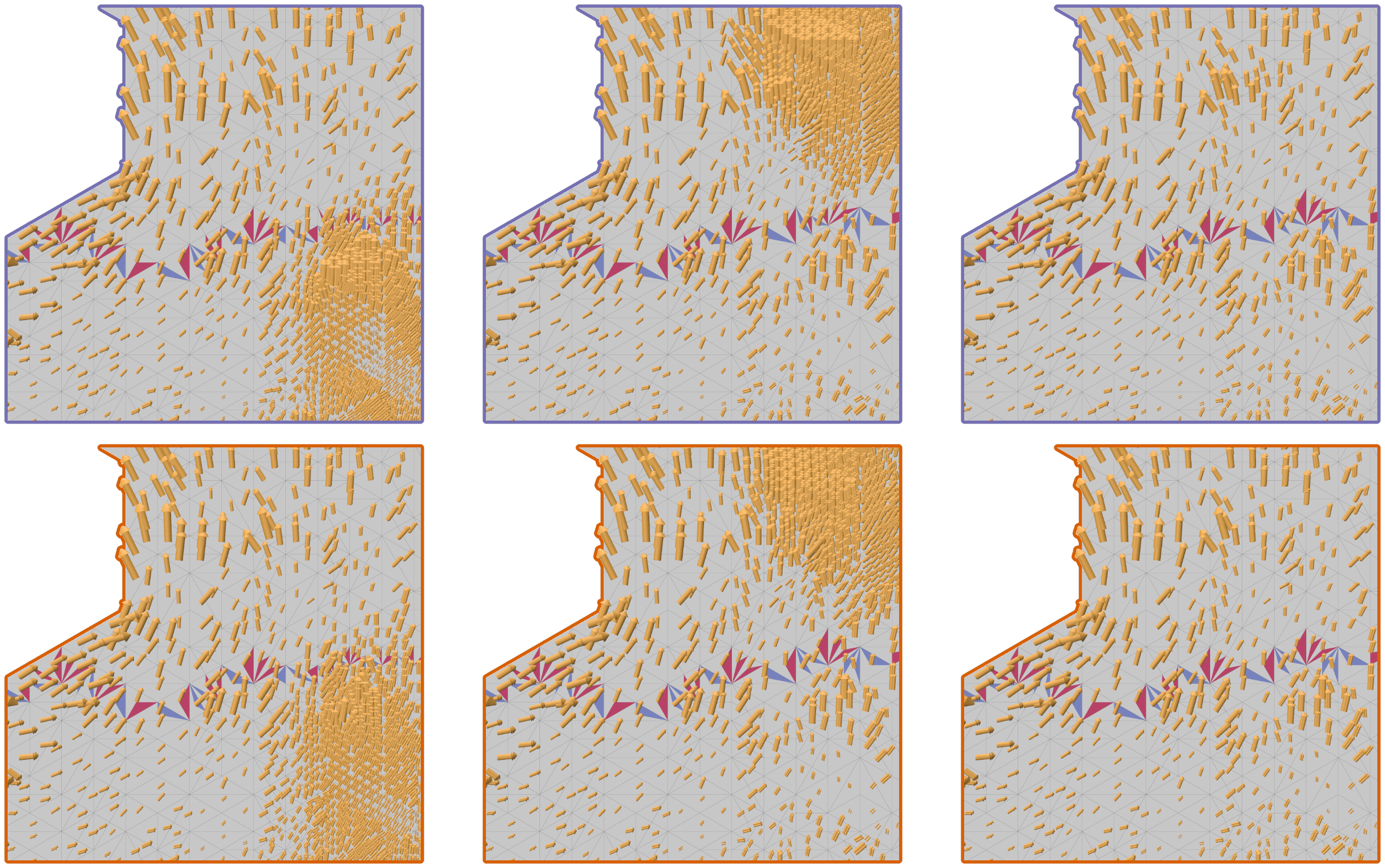}
    \caption{One harmonic basis vector field after $1$, $2$, and $3$ seconds of the simulation on the irregular mesh given in \Cref{figure:evaluation_irregular_mesh} and the path corresponding to this vector field.
    The interpolated vector field (top) exhibits artifacts where the vector of a triangle is copied to its children, whereas the recomputed vector field (bottom) yields a smoothly varying field.}
    \label{figure:evaluation_harmonic_basis_1}
\end{figure}

We also analyze the harmonic basis itself during the remeshing for both adaptive algorithms.
\Cref{figure:evaluation_harmonic_basis_1} displays one harmonic vector field after $1$, $2$, and $3$ seconds of the simulation.
Interpolating the vector field (top) results in a pattern of piecewise constant vectors as all refined faces obtain the same vector as their parents.
These artifacts are visible in the dense regions in \Cref{figure:evaluation_harmonic_basis_1}.
Such inaccuracies in the harmonic component of the fluid flow can mix with the curl-free component and cause further errors in the residual estimate $\varepsilon_R$ and the integrated vorticity.
A recomputation of the vector field (bottom) eliminates these artifacts, since the harmonicity and orthonormality guarantee the correct basis vector field.

\subsection{Vortex-Formation on a Torus}

We also perform a fluid simulation on the curved surface of a torus in \Cref{figure:teaser}.
In this case, we employ stricter error tolerances compared to the simulation on the discs and the exact values are given in a second example in \Cref{section:extended_evaluation} or in the implementation's repository.
The flow is initialized with a ring of negative vorticity on the top and a ring of positive vorticity on the bottom of the torus.
Moreover, the mesh is initially refined to have high resolution near the initial vorticity rings.
However, the simulation intentionally starts to break the vortex ring into several individual vortices.
While we again start with perfectly symmetric conditions, numerical errors quickly initiate chaotic behavior on the movement of the forming vortices such that they drift over the whole surface.
The static method is neither capable of capturing any dynamic details in the low-resolution areas nor of remaining stable for more than a few seconds.

Our adaptive algorithm using the recomputed harmonic basis experiences the chaotic vortex movements as well but remains stable in these situations.
Furthermore, we resolve highly detailed features of the vortices like their spiral arms and capture effects like merging vortices (see supplementary video).
In this setup, the initially refined faces are coarsened again when the residual error becomes low such that the number of triangles stays about constant during the simulation.

\section{Conclusion}

\subsection{Discussion}

In this paper, we introduced Adaptive Fluid Cohomology, a novel framework that successfully incorporates spatial and temporal adaptivity into the simulation of inviscid fluids on non-simply-connected curved surfaces.
By leveraging a posteriori error estimation and the Dormand-Prince 5(4) method, our approach dynamically targets computational resources to regions of high vorticity and complex flow evolution.
A pivotal component of our framework is the introduction of a topology-aware interpolation method for the harmonic basis vector fields.
Unlike naive $L^2$-projection strategies that disrupt the continuous homology class of the flow and induce numerical instability, our method robustly transfers harmonic forms across mutating meshes.

Our evaluations demonstrate that this adaptive approach yields significant computational advantages especially for computer graphics applications.
By dynamically coarsening and refining the mesh, we achieved up to an $86\,\%$ reduction in memory footprint without sacrificing the visual or physical fidelity of the dynamics compared to high-resolution static simulations.
Furthermore, our algorithm exhibits superior robustness, maintaining stable and physically accurate flows even on poor-quality, irregular triangulations where traditional static solvers fail.
Our results indicate that the computational overhead introduced by the adaptation framework is effectively offset by the dynamic temporal resolution, allowing the adaptive simulations to achieve runtimes comparable to, or even lower than, those of the static approach.

\subsection{Future Work}

Our framework presents an advancement for surface-based fluid simulation.
However, several exciting avenues remain for future research.
While the core theoretical foundations of our adaptive fluid cohomology framework are inherently generalizable to three-dimensional domains, several aspects of our current implementation, e.g. the handling of intrinsic triangulations, are explicitly tailored to the structure of 2D surfaces.
Although extensions of the NVB algorithm to higher dimensions already exist \cite{stevenson2008completion}, we need an efficient data structure that also supports a corresponding coarsening operation.
It will be particularly compelling to evaluate the computational benefits in 3D settings, where the potential for memory savings and runtime efficiency may prove even more significant given the increased complexity of volumetric discretization.
Additionally, our present focus is on idealized fluids governed by the Euler equations on fixed domains.
Incorporating viscosity to simulate Navier-Stokes flows or dynamic domains would broaden the applicability of our method to a wider range of natural phenomena.
The introduction of other remeshing algorithms potentially paired with a different interpolation scheme would be a promising first step towards these goals.

Finally, exploring hardware acceleration and optimized data structures for dynamic intrinsic triangulations could further reduce runtime overhead, enabling the real-time simulation for large-scale interactive graphics and visual effects.
We believe that the development of a sparse linear system library specifically for adaptive simplicial complexes would be highly beneficial.
Also, methodological changes to an implicit time integration scheme usually make simulations even more robust and may have little computational overhead when combined with the mentioned hardware and data structure optimizations.

\subsection*{Acknowledgements}

This work has been funded by the Ministry of Culture and Science North Rhine-Westphalia under grant number PB22-063A (InVirtuo 4.0: Experimental Research in Virtual Environments) and by the Federal Ministry of Education and Research under Grant No. 01IS22094A WEST-AI.
\appendix
\section{Extended Evaluation}
\label{section:extended_evaluation}

Here, we provide additional experiments on high-genus meshes under analogous conditions.

\subsection{High-Genus Disc}

\begin{figure}[t]
    \centering
    \includegraphics[width=0.999\linewidth]{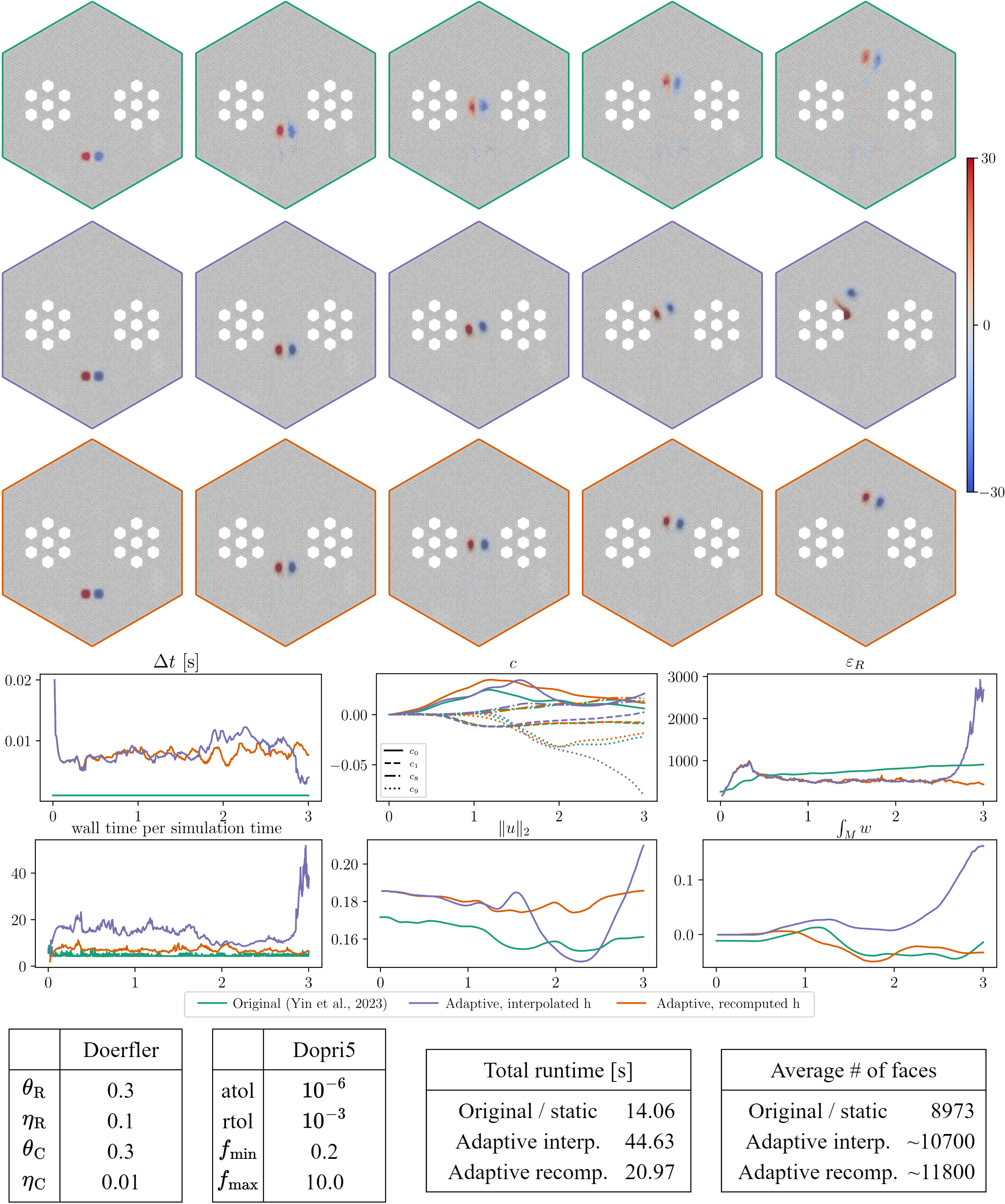}
    \caption{Simulation of a vortex pair on an irregular hexagonal disc with $14$ holes.
        While the static baseline (top, green border) and our adaptive method with recomputed harmonic basis fields (bottom, orange border) exhibit a slight rightward drift, the interpolated harmonic basis fields (center, blue border) cause the vortices to drift leftward into a hole, ultimately leading to simulation failure.}
    \label{figure:evaluation_14_holes_irregular}
\end{figure}

First, we modify the irregular hexagonal disc to contain a total of $14$ holes instead of two.
As a result, the fluid flow contains $14$ time-varying harmonic components.
To improve readability in the corresponding diagram, we only report the harmonic coefficients $c_0, c_1, c_8$, and $c_9$.

The results for the static baseline as well as the adaptive approaches using interpolated and recomputed basis fields are shown in \Cref{figure:evaluation_14_holes_irregular}.
In this experiment, we restrict the comparison to the first three seconds of simulated time because the adaptive method with interpolated basis fields causes the vortices to drift into a hole, leading to a simulation failure.
The most noticeable difference between the three simulations is the direction of vortex drift.
While the vortices in the static baseline and in our adaptive method with recomputed basis fields drift slightly to the right, the vortices in the simulation using interpolated basis fields drift to the left.
Consequently, significant deviations emerge toward the end of the simulation in several harmonic coefficients $c_i$, the average flow velocity $\Vert u \Vert_2$, and the error metrics $\varepsilon_R$ and $\int_M \omega$.
Despite these discrepancies becoming apparent only later in the simulation, the interpolated approach exhibits substantially worse runtime performance from the beginning, even when compared to the other adaptive method.

Comparing our adaptive approach with recomputed basis fields to the static baseline reveals only minor differences in the simulation results.
Despite operating with a considerably smaller time step size, the baseline method remains approximately $33\,\%$ faster than our adaptive solver.
This performance gap is primarily caused by the increased cost associated with handling the harmonic components.
For the case with two holes, updating the harmonic basis accounted for only about $10\,\%$ of the total computation time.
With $14$ holes, however, this share rises to approximately $34\,\%$.
Nevertheless, this limitation can be mitigated through parallelization, since the harmonic components can be processed independently and concurrently up to the orthonormalization step.

\subsection{High-Genus Torus}

As a second example, we perform a simulation on a torus augmented with four cylindrical tubes that extend radially inward toward the central cavity.
This geometry gives rise to a total of $8$ harmonic components.

As in the regular torus experiment presented in the main paper, the static method fails to accurately resolve vortices once they leave the initially refined region.
Consequently, the simulation becomes unstable and terminates after approximately three seconds, despite employing very small time steps of only $\Delta t = 10^{-4}\,\mathrm{s}$.
As a result, the overall computation time is roughly $50$ times longer than the simulated physical time.
In contrast, our adaptive approach with recomputed harmonic basis fields successfully captures the evolving flow structures throughout the simulation.
Although we employ stricter error tolerances than in the disc experiments, the simulation completes in less than four times the simulated time.
Owing to the adaptive spatial refinement, time step sizes of $\Delta t = 0.01\,\mathrm{s}$ can be used while maintaining sufficient accuracy.

\begin{figure}[t]
    \centering
    \includegraphics[width=0.999\linewidth]{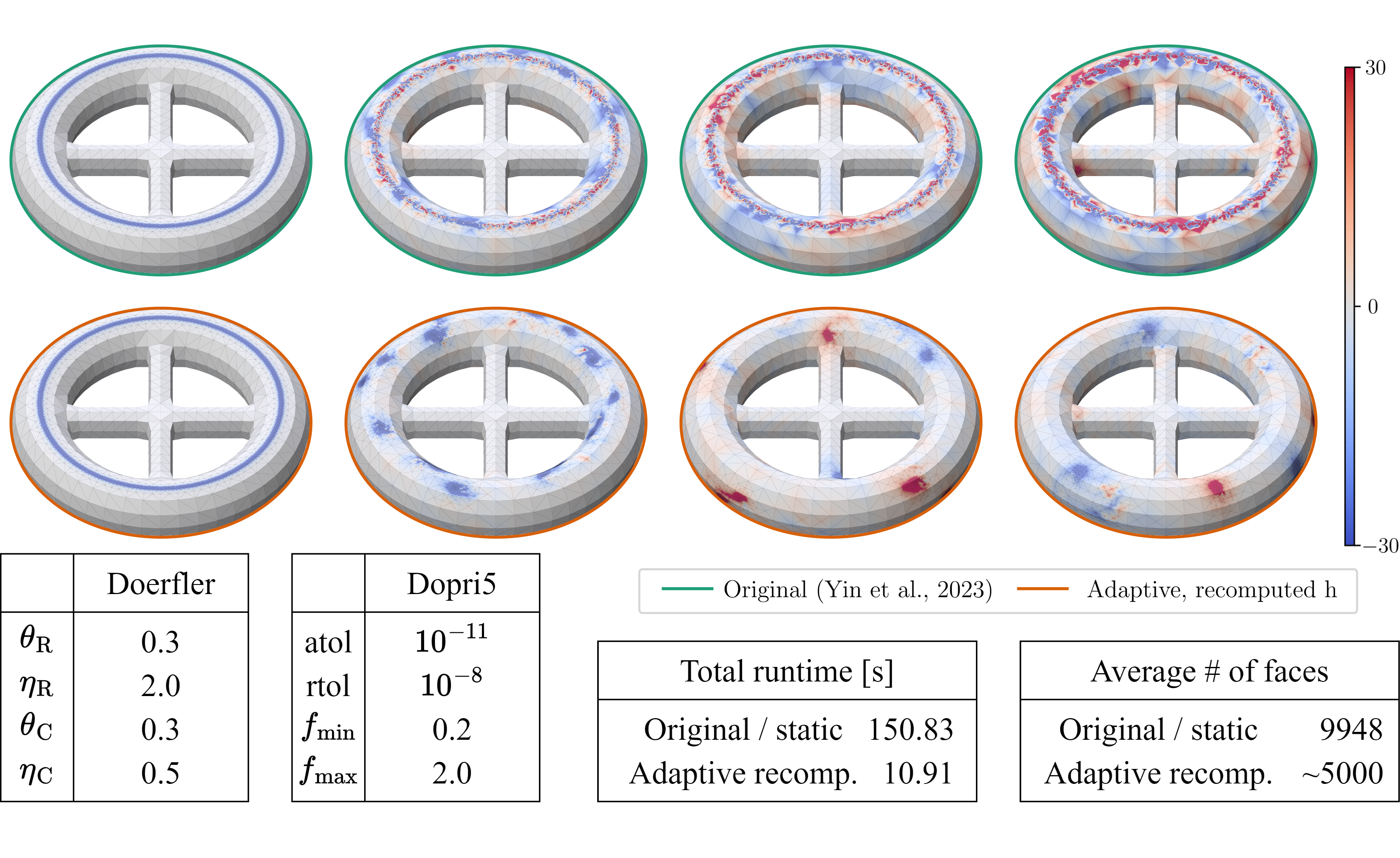}
    \caption{Simulation on a torus with $4$ inwards facing cylindrical tubes.
        Similar to the regular torus, the static method fails quickly due to different triangle sizes even with small time steps of $\Delta t = 10^{-4}\,\mathrm{s}$.
        In contrast, our adaptive method simulates complex flow patterns emerging on the surface.}
    \label{figure:evaluation_torus_4}
\end{figure}

\section{Implementation}
\label{section:implementation}

Here, we provide the subroutines used by the "Adapt Spatial Domain" algorithm, along with additional procedures referenced throughout the main paper. References to sections mentioned in the accompanying comments correspond to the respective sections in the main paper.

\begin{algorithm}[ht]
    \caption{Delta Form}
    \label{algorithm:delta_form}
    \KwData{Cycle $\bm{\sigma_i}$ (e.g. list of edges that are crossed by $\bm{\sigma}$)}
    \KwResult{1-form $\eta_i$ with $\mathrm{d} \eta_i = 0$}

    Initialize $\eta_i \in \Omega^1 \gets 0$\;
    \For{edge $e \in \bm{\sigma}$}
    {
        \tcp{Set $\eta_i$ such that $\mathrm{d} \eta_i = 0$}
        \If{dual edge of $\bm{\sigma}$ crosses $e$ from left to right}
        {
            $\eta_i(e) \gets 1$\;
        }
        \Else
        {
            $\eta_i(e) \gets -1$\;
        }
    }
    \Return{$\eta_i$}\;
\end{algorithm}

\begin{algorithm}[ht]
    \caption{Harmonic Basis}
    \label{algorithm:harmonic_basis}
    \KwData{Cycles $\bm{\sigma}_1, \dots, \bm{\sigma}_m$}
    \KwResult{1-form $\eta_i$ with $\mathrm{d} \eta_i = 0$}

    Initialize $\bm{h} = 0 \in \Omega^1 \gets 0$\;
    \For{$i = 1$ to $m$}
    {
        $\eta_i \gets$ \textsc{DeltaForm}($\bm{\sigma}_i$) \tcp*{\Cref{algorithm:delta_form}}
        $h_i \gets \mathrm{proj}_{\mathrm{ker}(\delta)}(\eta_i)$ \tcp*{Section 3.2.3}
        $h_i \gets$ \textsc{WhitneyInterpolation}($h_i$)\;
    }
    \Return{$\bm{h}$}\;
\end{algorithm}

\begin{algorithm}[ht]
    \caption{Bisect Edge}
    \label{algorithm:bisect_edge}
    \KwData{intrinsic triangulation $\mathcal{T}$ of mesh $M = (V, E, F)$, edge $e \in E$, indexing function $i \in (F \to \mathbb{Z})$}

    $\textsc{SplitEdge}(\mathcal{T}, e, 0.5)$ \tcp*{\cite{Gillespie2021intrinsic}}
    \For{side $(l, r) \in F^2$ in $\textsc{Neighbor}(\text{old edge } e)$}
    {
        $\mathbf{if}$ $i(l) > i(r)$: swap $i(l)$ and $i(r)$\;
    }
\end{algorithm}

\begin{algorithm}[ht]
    \caption{Refine}
    \label{algorithm:refine}
    \KwData{intrinsic triangulation $\mathcal{T}$ of mesh $M = (V, E, F)$, marked faces $S = \mathrm{Stack}(F)$, refinement edges $r : F \to E$}
    \KwResult{Refined mesh}

    $E_0 \gets \mathrm{Stack}(E)$\;
    \tcp{Iteratively mark all faces affected}
    \While{$S \neq \emptyset$}
    {
        $f \gets \textsc{Pop}(S)$, mark $f$\;
        $\mathbf{if}$ $\textsc{Neighbor}(f)$ not marked: \textsc{Push} $\textsc{Neighbor}(f)$ onto $S$\;
        $\mathbf{if}$ $r(f)$ is refinable: \textsc{Push} $r(f)$ onto $E_0$\;
    }
    \tcp{Do refinement}
    \While{$E_0 \neq \emptyset$}
    {
        $e \gets \textsc{Pop}(E_0)$, unmark neighboring faces\;
        $v \gets \textsc{BisectEdge}(\mathcal{T}, e)$ \tcp*{\Cref{algorithm:bisect_edge}}
        \tcp{Repeat iteratively if possible}
        \For{face $f$ adjacent to $v$}
        {
            $\mathbf{if}$ $\textsc{Neighbor}(f)$ is marked and $r(f)$ if refineable: \textsc{Push} $r(f)$ onto $E_0$\;
        }
    }
\end{algorithm}

\begin{algorithm}[ht]
    \caption{Find Parent Edge}
    \label{algorithm:find_parent_edge}
    \KwData{vertex $v \in V$, indexing function $i \in (F \to \mathbb{Z})$}

    $\vec{vj} \gets$ outgoing halfedge of $v$ with minimal face index $i$\;
    \Return{\textsc{Next}(\textsc{Twin}($\vec{vj}$))}
\end{algorithm}

\begin{algorithm}[ht]
    \caption{Collapse Edge}
    \label{algorithm:collapse_edge}
    \KwData{intrinsic triangulation $\mathcal{T} = (M, M_i, l, n, r)$ of mesh $M = (V, E, F)$, halfedge $\vec{vj}$}

    $e \in E \gets \textsc{CollapseEdgeTriangular}(\mathcal{T}, \vec{vj})$ \tcp*{\cite{Sharp2019intrinsic}}
    \If{$n_{ip} = -1 \lor n_{pj} = -1$}
    {
        $n_{ij} \gets -1$\;
    }
    \Else
    {
        $n_{ij} = n_{ip} + n_{pj}$
    }
    $r_{\vec{ij}} \gets r_{\vec{vj}}$\;
    $r_{\vec{ji}} \gets r_{\vec{iv}}$\;
    $i_{ij} \gets l_{ip} + l_{pj}$\;
\end{algorithm}

\begin{algorithm}[ht]
    \caption{Unify Vertex}
    \label{algorithm:unify_vertex}
    \KwData{intrinsic triangulation $\mathcal{T}$ of mesh $M = (V, E, F)$, vertex $v \in V$, indexing function $i \in (F \to \mathbb{Z})$}
    \KwResult{Merged edge $e$}

    $\vec{vj} = \textsc{FindParentEdge}(v, i)$ \tcp*{\Cref{algorithm:find_parent_edge}}
    $e \in E \gets \textsc{CollapseEdge}(\mathcal{T}, \vec{vj}, 0.5)$ \tcp*{\Cref{algorithm:collapse_edge}}
    \For{neighboring face $f$ of $e$}
    {
        $i(f) \gets$ index of left face in old triangulation\;
    }
    \Return{$e$}
\end{algorithm}

\begin{algorithm}[ht]
    \caption{Coarsen}
    \label{algorithm:coarsen}
    \KwData{intrinsic triangulation $\mathcal{T}$ of mesh $M = (V, E, F)$, marked faces $S = \mathrm{Stack}(F)$, refinement edges $r : F \to E$}
    \KwResult{Coarsened mesh}

    $V_0 \gets \mathrm{Stack}(E)$\;
    \tcp{Iteratively mark all faces affected}
    \While{$S \neq \emptyset$}
    {
        $f \gets \textsc{Pop}(S)$, mark $f$\;
        \ForEach{vertex $v$ of $f$}
        {
            \If{$v$ is coarsenable and all adjacent faces are marked}
            {
                \textsc{Push} $v$ onto $V_0$\;
            }
        }
    }
    \tcp{Do coarsening}
    \While{$V_0 \neq \emptyset$}
    {
        $v \gets \textsc{Pop}(V_0)$, unmark neighboring faces\;
        $e \gets \textsc{UnifyVertex}(\mathcal{T}, v)$ \tcp*{\Cref{algorithm:unify_vertex}}
        \tcp{Repeat iteratively if possible}
        \ForEach{vertex $u$ opposite to $e$}
        {
            \If{$u$ is coarsenable and all its adjacent faces are marked}
            {
                \textsc{Push} $u$ onto $V_0$\;
            }
        }
    }
\end{algorithm}

\clearpage

%-------------------------------------------------------------------------
% bibtex
\bibliographystyle{eg-alpha-doi}
\bibliography{egbibsample}

@article{yin_2023_fluid_cohomology,
    author = {Yin, Hang and Nabizadeh, Mohammad Sina and Wu, Baichuan and Wang, Stephanie and Chern, Albert},
    title = {Fluid Cohomology},
    year = {2023},
    publisher = {Association for Computing Machinery},
    volume = {42},
    number = {4},
    issn = {0730-0301},
    doi = {10.1145/3592402},
    journal = {ACM Trans. Graph.},
    numpages = {25}
}

@article{Nabizadeh_2022_covector_fluids,
    author = {Nabizadeh, Mohammad Sina and Wang, Stephanie and Ramamoorthi, Ravi and Chern, Albert},
    title = {Covector fluids},
    year = {2022},
    issue_date = {July 2022},
    publisher = {Association for Computing Machinery},
    volume = {41},
    number = {4},
    issn = {0730-0301},
    doi = {10.1145/3528223.3530120},
    journal = {ACM Trans. Graph.},
    articleno = {113},
    numpages = {16},
}

@inproceedings{stam_1999_fluids,
    author = {Stam, Jos},
    title = {Stable fluids},
    year = {1999},
    publisher = {ACM Press/Addison-Wesley Publishing Co.},
    doi = {10.1145/311535.311548},
    booktitle = {Proceedings of the 26th Annual Conference on Computer Graphics and Interactive Techniques},
    pages = {121–128},
    numpages = {8},
    series = {SIGGRAPH '99}
}

@inproceedings{foster1997modeling,
    title={Modeling the motion of a hot, turbulent gas},
    author={Foster, Nick and Metaxas, Dimitris},
    booktitle={Proceedings of the 24th annual conference on Computer graphics and interactive techniques},
    pages={181--188},
    year={1997}
}

@book{forsythe1965finite,
    title={Finite-difference methods for partial differential equations},
    author={Forsythe, George E and Wasow, Wolfgang Richard},
    volume={196},
    year={1965},
    publisher={Wiley New York}
}

@techreport{harlow1962particle,
    title={The particle-in-cell method for numerical solution of problems in fluid dynamics},
    author={Harlow, Francis H},
    year={1962},
    institution={Los Alamos National Laboratory (LANL), Los Alamos, NM (United States)}
}

@article{harlow1965numerical,
    title={Numerical calculation of time-dependent viscous incompressible flow of fluid with free surface},
    author={Harlow, Francis H and Welch, J Eddie and others},
    journal={Physics of fluids},
    volume={8},
    number={12},
    pages={2182},
    year={1965}
}

@article{Batty2007_solid_fluid,
    author = {Batty, Christopher and Bertails, Florence and Bridson, Robert},
    title = {A fast variational framework for accurate solid-fluid coupling},
    year = {2007},
    publisher = {Association for Computing Machinery},
    volume = {26},
    number = {3},
    issn = {0730-0301},
    doi = {10.1145/1276377.1276502},
    journal = {ACM Trans. Graph.},
    pages = {100–es},
    numpages = {8},
}

@article{LAI_1994_FVM,
    author = {Y. G. LAI and A. J. PRZEKWAS},
    title = {A FINITE-VOLUME METHOD FOR FLUID FLOW SIMULATIONS WITH MOVING BOUNDARIES},
    journal = {International Journal of Computational Fluid Dynamics},
    volume = {2},
    number = {1},
    pages = {19--40},
    year = {1994},
    publisher = {IAHR Website},
    doi = {10.1080/10618569408904482},
}

@Inbook{Moukalled2016,
    author="Moukalled, F. and Mangani, L. and Darwish, M.",
    title="The Finite Volume Method",
    bookTitle="The Finite Volume Method in Computational Fluid Dynamics: An Advanced Introduction with OpenFOAM® and Matlab",
    year="2016",
    publisher="Springer International Publishing",
    address="Cham",
    pages="103--135",
    isbn="978-3-319-16874-6",
    doi="10.1007/978-3-319-16874-6_5",
}

@article{Dalal07082008,
    author = {Amaresh Dalal and V. Eswaran and G. Biswas},
    title = {A Finite-Volume Method for Navier-Stokes Equations on Unstructured Meshes},
    journal = {Numerical Heat Transfer, Part B: Fundamentals},
    volume = {54},
    number = {3},
    pages = {238--259},
    year = {2008},
    publisher = {Taylor \& Francis},
    doi = {10.1080/10407790802182653},
}

@article{BRACKBILL198825,
    title = {Flip: A low-dissipation, particle-in-cell method for fluid flow},
    author = {J.U. Brackbill and D.B. Kothe and H.M. Ruppel},
    journal = {Computer Physics Communications},
    volume = {48},
    number = {1},
    pages = {25-38},
    year = {1988},
    issn = {0010-4655},
    doi = {https://doi.org/10.1016/0010-4655(88)90020-3},
}

@article{Yongning_2005_sand,
    author = {Zhu, Yongning and Bridson, Robert},
    title = {Animating sand as a fluid},
    year = {2005},
    publisher = {Association for Computing Machinery},
    address = {New York, NY, USA},
    volume = {24},
    number = {3},
    issn = {0730-0301},
    doi = {10.1145/1073204.1073298},
    journal = {ACM Trans. Graph.},
    pages = {965–972},
    numpages = {8},
}

@inproceedings{Ando_2011_sheets,
    author = {Ando, Ryoichi and Tsuruno, Reiji},
    title = {A particle-based method for preserving fluid sheets},
    year = {2011},
    isbn = {9781450309233},
    publisher = {Association for Computing Machinery},
    doi = {10.1145/2019406.2019408},
    booktitle = {Proceedings of the 2011 ACM SIGGRAPH/Eurographics Symposium on Computer Animation},
    pages = {7–16},
    numpages = {10},
    series = {SCA '11}
}

@article{Monaghan_2005,
    doi = {10.1088/0034-4885/68/8/R01},
    year = {2005},
    month = {jul},
    volume = {68},
    number = {8},
    pages = {1703},
    author = {Monaghan, J J},
    title = {Smoothed particle hydrodynamics},
    journal = {Reports on Progress in Physics},
}

@article{PRICE2012759,
    title = {Smoothed particle hydrodynamics and magnetohydrodynamics},
    author = {Daniel J. Price},
    journal = {Journal of Computational Physics},
    volume = {231},
    number = {3},
    pages = {759-794},
    year = {2012},
    note = {Special Issue: Computational Plasma Physics},
    issn = {0021-9991},
    doi = {https://doi.org/10.1016/j.jcp.2010.12.011},
}

@article{koschier2020smoothed,
    title={Smoothed particle hydrodynamics techniques for the physics based simulation of fluids and solids},
    author={Koschier, Dan and Bender, Jan and Solenthaler, Barbara and Teschner, Matthias},
    journal={arXiv preprint arXiv:2009.06944},
    year={2020}
}

@inproceedings{becker2007weakly,
    title={Weakly compressible SPH for free surface flows},
    author={Becker, Markus and Teschner, Matthias},
    booktitle={Proceedings of the 2007 ACM SIGGRAPH/Eurographics symposium on Computer animation},
    pages={209--217},
    year={2007}
}

@inproceedings{Solenthaler_2009_incompressible_sph,
    author = {Solenthaler, B. and Pajarola, R.},
    title = {Predictive-corrective incompressible SPH},
    year = {2009},
    isbn = {9781605587264},
    publisher = {Association for Computing Machinery},
    doi = {10.1145/1576246.1531346},
    booktitle = {ACM SIGGRAPH 2009 Papers},
    articleno = {40},
    numpages = {6},
    series = {SIGGRAPH '09}
}

@article{kelager2006lagrangian,
    title={Lagrangian fluid dynamics using smoothed particle hydrodynamics},
    author={Kelager, Micky},
    journal={University of Copenhagen: Department of Computer Science},
    volume={2},
    year={2006}
}

@article{SULSKY1995236,
    title = {Application of a particle-in-cell method to solid mechanics},
    author = {Deborah Sulsky and Shi-Jian Zhou and Howard L. Schreyer},
    journal = {Computer Physics Communications},
    volume = {87},
    number = {1},
    pages = {236-252},
    year = {1995},
    issn = {0010-4655},
    doi = {https://doi.org/10.1016/0010-4655(94)00170-7},
}

@article{Fei_2021_MPM,
    author = {Fei, Yun (Raymond) and Guo, Qi and Wu, Rundong and Huang, Li and Gao, Ming},
    title = {Revisiting integration in the material point method: a scheme for easier separation and less dissipation},
    year = {2021},
    publisher = {Association for Computing Machinery},
    volume = {40},
    number = {4},
    issn = {0730-0301},
    doi = {10.1145/3450626.3459678},
    journal = {ACM Trans. Graph.},
    articleno = {109},
    numpages = {16},
}

@article{Jiang_affine_PIC,
    author = {Jiang, Chenfanfu and Schroeder, Craig and Selle, Andrew and Teran, Joseph and Stomakhin, Alexey},
    title = {The affine particle-in-cell method},
    year = {2015},
    publisher = {Association for Computing Machinery},
    volume = {34},
    number = {4},
    issn = {0730-0301},
    doi = {10.1145/2766996},
    journal = {ACM Trans. Graph.},
    articleno = {51},
    numpages = {10},
}

@article{Stomakhin_2013_snow,
    author = {Stomakhin, Alexey and Schroeder, Craig and Chai, Lawrence and Teran, Joseph and Selle, Andrew},
    title = {A material point method for snow simulation},
    year = {2013},
    issue_date = {July 2013},
    publisher = {Association for Computing Machinery},
    volume = {32},
    number = {4},
    issn = {0730-0301},
    doi = {10.1145/2461912.2461948},
    journal = {ACM Trans. Graph.},
    articleno = {102},
    numpages = {10},
}

@article{Chen_1998_LBM,
    author = {Chen, Shiyi and Doolen, Gary D.},
    title = {LATTICE BOLTZMANN METHOD FOR FLUID FLOWS},
    journal= {Annual Review of Fluid Mechanics},
    year = {1998},
    volume = {30},
    pages = {329--364},
    doi = {https://doi.org/10.1146/annurev.fluid.30.1.329},
    publisher = {Annual Reviews},
}

@article{Aidun_2021_LBMcomplex,
    author = {Aidun, Cyrus K. and Clausen, Jonathan R.},
    title = {Lattice-Boltzmann Method for Complex Flows},
    journal= {Annual Review of Fluid Mechanics},
    year = {2010},
    volume = {42},
    pages = {439--472},
    doi = {https://doi.org/10.1146/annurev-fluid-121108-145519},
    publisher = {Annual Reviews},
}

@article{Li_2016_LBM,
    title = {Lattice Boltzmann methods for multiphase flow and phase-change heat transfer},
    author = {Q. Li and K.H. Luo and Q.J. Kang and Y.L. He and Q. Chen and Q. Liu},
    journal = {Progress in Energy and Combustion Science},
    volume = {52},
    pages = {62--105},
    year = {2016},
    doi = {https://doi.org/10.1016/j.pecs.2015.10.001},
}

@article{huang2015multiphase,
    title = {Multiphase lattice Boltzmann methods: Theory and application},
    author = {Huang, Haibo and Sukop, Michael and Lu, Xiyun},
    year = {2015},
    publisher = {John Wiley \& Sons}
}

@book{guo2013lattice,
    title={Lattice Boltzmann method and its application in engineering},
    author={Guo, Zhaoli and Shu, Chang},
    volume={3},
    year={2013},
    publisher={World Scientific}
}

@InProceedings{Tompson_2017_accelerating,
    title = {Accelerating {E}ulerian Fluid Simulation With Convolutional Networks},
    author = {Jonathan Tompson and Kristofer Schlachter and Pablo Sprechmann and Ken Perlin},
    booktitle = {Proceedings of the 34th International Conference on Machine Learning},
    pages = {3424--3433},
    year = {2017},
    editor = {Precup, Doina and Teh, Yee Whye},
    volume = {70},
    series = {Proceedings of Machine Learning Research},
    month = {06--11 Aug},
    publisher = {PMLR},
    url = {https://proceedings.mlr.press/v70/tompson17a.html},
}

@inproceedings{Dong_2019_nn,
    author = {Dong, Wenqian and Liu, Jie and Xie, Zhen and Li, Dong},
    title = {Adaptive neural network-based approximation to accelerate eulerian fluid simulation},
    year = {2019},
    isbn = {9781450362290},
    publisher = {Association for Computing Machinery},
    doi = {10.1145/3295500.3356147},
    booktitle = {Proceedings of the International Conference for High Performance Computing, Networking, Storage and Analysis},
    articleno = {7},
    numpages = {22},
    location = {Denver, Colorado},
    series = {SC '19}
}

@InProceedings{Sanchez_Gonzalez_2020_complex,
    title = {Learning to Simulate Complex Physics with Graph Networks},
    author = {Sanchez-Gonzalez, Alvaro and Godwin, Jonathan and Pfaff, Tobias and Ying, Rex and Leskovec, Jure and Battaglia, Peter},
    booktitle = {Proceedings of the 37th International Conference on Machine Learning},
    pages = {8459--8468},
    year = {2020},
    editor = {III, Hal Daumé and Singh, Aarti},
    volume = {119},
    series = {Proceedings of Machine Learning Research},
    month = {13--18 Jul},
    publisher = {PMLR},
    url = {https://proceedings.mlr.press/v119/sanchez-gonzalez20a.html},
}

@inproceedings{wandel2021fluid,
    title = {Learning {Incompressible} {Fluid} {Dynamics} from {Scratch} - {Towards} {Fast}, {Differentiable} {Fluid} {Models} that {Generalize}},
    author = {Wandel, Nils and Weinmann, Michael and Klein, Reinhard},
    booktitle = {International {Conference} on {Learning} {Representations} ({ICLR})},
    year = {2021},
}

@ARTICLE{wang_2024_survey,
    author={Wang, Xiaokun and Xu, Yanrui and Liu, Sinuo and Ren, Bo and Kosinka, Jirí and Telea, Alexandru C. and Wang, Jiamin and Song, Chongming and Chang, Jian and Li, Chenfeng and Zhang, Jian Jun and Ban, Xiaojuan},
    journal={Computational Visual Media},
    title={Physics-based fluid simulation in computer graphics: Survey, research trends, and challenges},
    year={2024},
    volume={10},
    number={5},
    pages={803-858},
    doi={10.1007/s41095-023-0368-y}
}

@inproceedings{Hu_2019_survey_MPM,
    author = {Hu, Yuanming and Zhang, Xinxin and Gao, Ming and Jiang, Chenfanfu},
    title = {On hybrid lagrangian-eulerian simulation methods: practical notes and high-performance aspects},
    year = {2019},
    isbn = {9781450363075},
    publisher = {Association for Computing Machinery},
    doi = {10.1145/3305366.3328075},
    booktitle = {ACM SIGGRAPH 2019 Courses},
    articleno = {16},
    numpages = {246},
    series = {SIGGRAPH '19}
}

@article{Fromm_1963_vorticity_streamfunction,
    author = {Fromm, Jacob E. and Harlow, Francis H.},
    title = {Numerical Solution of the Problem of Vortex Street Development},
    journal = {The Physics of Fluids},
    volume = {6},
    number = {7},
    pages = {975-982},
    year = {1963},
    month = {07},
    issn = {0031-9171},
    doi = {10.1063/1.1706854},
}

@article{Cheng1972numerical,
    author = {Cheng, Ralph Ta‐shun},
    title = {Numerical Solution of the Navier‐Stokes Equations by the Finite Element Method},
    journal = {The Physics of Fluids},
    volume = {15},
    number = {12},
    pages = {2098-2105},
    year = {1972},
    month = {12},
    issn = {0031-9171},
    doi = {10.1063/1.1693841},
}

@article{Elcott2007simplicial,
    author = {Elcott, Sharif and Tong, Yiying and Kanso, Eva and Schr\"{o}der, Peter and Desbrun, Mathieu},
    title = {Stable, circulation-preserving, simplicial fluids},
    year = {2007},
    issue_date = {January 2007},
    publisher = {Association for Computing Machinery},
    volume = {26},
    number = {1},
    issn = {0730-0301},
    doi = {10.1145/1189762.1189766},
    journal = {ACM Trans. Graph.},
    pages = {4–es},
    numpages = {12},
}

@inproceedings{selle2005vortex,
    author = {Selle, Andrew and Rasmussen, Nick and Fedkiw, Ronald},
    title = {A vortex particle method for smoke, water and explosions},
    year = {2005},
    isbn = {9781450378253},
    publisher = {Association for Computing Machinery},
    doi = {10.1145/1186822.1073282},
    booktitle = {ACM SIGGRAPH 2005 Papers},
    pages = {910–914},
    numpages = {5},
    location = {Los Angeles, California},
    series = {SIGGRAPH '05},
}

@inproceedings{meldgaard2022fast,
    title={Fast Vortex Particle Method for Fluid-Character Interaction},
    author={Asger Meldgaard and Sune Darkner and Kenny Erleben},
    booktitle={Graphics Interface 2022},
    year={2022},
}

@inproceedings{weissmann2010filament,
    author = {Weißmann, Steffen and Pinkall, Ulrich},
    title = {Filament-based smoke with vortex shedding and variational reconnection},
    year = {2010},
    isbn = {9781450302104},
    publisher = {Association for Computing Machinery},
    doi = {10.1145/1833349.1778852},
    booktitle = {ACM SIGGRAPH 2010 Papers},
    articleno = {115},
    numpages = {12},
    location = {Los Angeles, California},
    series = {SIGGRAPH '10},
}

@inproceedings{barnat2012smoke,
    title={Smoke sheets for graph-structured vortex filaments},
    author={Barnat, Alfred and Pollard, Nancy S},
    booktitle={Proceedings of the 11th ACM SIGGRAPH/Eurographics conference on Computer Animation},
    pages={77--86},
    year={2012}
}

@article{Pfaff2012sheets,
    author = {Pfaff, Tobias and Thuerey, Nils and Gross, Markus},
    title = {Lagrangian vortex sheets for animating fluids},
    year = {2012},
    issue_date = {July 2012},
    publisher = {Association for Computing Machinery},
    volume = {31},
    number = {4},
    issn = {0730-0301},
    doi = {10.1145/2185520.2185608},
    journal = {ACM Trans. Graph.},
    month = jul,
    articleno = {112},
    numpages = {8},
}

@article{Golas2012largescale,
    author = {Golas, Abhinav and Narain, Rahul and Sewall, Jason and Krajcevski, Pavel and Dubey, Pradeep and Lin, Ming},
    title = {Large-scale fluid simulation using velocity-vorticity domain decomposition},
    year = {2012},
    issue_date = {November 2012},
    publisher = {Association for Computing Machinery},
    volume = {31},
    number = {6},
    issn = {0730-0301},
    doi = {10.1145/2366145.2366167},
    journal = {ACM Trans. Graph.},
    month = nov,
    articleno = {148},
    numpages = {9},
}

@inproceedings{Park2005hybrid,
    author = {Park, Sang Il and Kim, Myoung Jun},
    title = {Vortex fluid for gaseous phenomena},
    year = {2005},
    isbn = {1595931988},
    publisher = {Association for Computing Machinery},
    doi = {10.1145/1073368.1073406},
    booktitle = {Proceedings of the 2005 ACM SIGGRAPH/Eurographics Symposium on Computer Animation},
    pages = {261–270},
    numpages = {10},
    location = {Los Angeles, California},
    series = {SCA '05}
}

@article{Cortez1996impulse,
    title = {An Impulse-Based Approximation of Fluid Motion due to Boundary Forces},
    author = {Ricardo Cortez},
    journal = {Journal of Computational Physics},
    volume = {123},
    number = {2},
    pages = {341-353},
    year = {1996},
    issn = {0021-9991},
    doi = {https://doi.org/10.1006/jcph.1996.0028},
}

@ARTICLE{feng2023impulse,
    author={Feng, Fan and Liu, Jinyuan and Xiong, Shiying and Yang, Shuqi and Zhang, Yaorui and Zhu, Bo},
    journal={IEEE Transactions on Visualization and Computer Graphics},
    title={Impulse Fluid Simulation},
    year={2023},
    volume={29},
    number={6},
    pages={3081-3092},
    doi={10.1109/TVCG.2022.3149466}
}

@inproceedings{deGoes2016fields,
    author = {de Goes, Fernando and Desbrun, Mathieu and Tong, Yiying},
    title = {Vector field processing on triangle meshes},
    year = {2016},
    isbn = {9781450342896},
    publisher = {Association for Computing Machinery},
    doi = {10.1145/2897826.2927303},
    booktitle = {ACM SIGGRAPH 2016 Courses},
    articleno = {27},
    numpages = {49},
    series = {SIGGRAPH '16}
}

@ARTICLE{Bhatia2013decomposition,
    author={Bhatia, H. and Norgard, G. and Pascucci, V. and Bremer, Peer-Timo},
    journal={IEEE Transactions on Visualization \& Computer Graphics },
    title={{The Helmholtz-Hodge Decomposition—A Survey}},
    year={2013},
    volume={19},
    number={08},
    ISSN={1941-0506},
    pages={1386-1404},
    doi={10.1109/TVCG.2012.316},
    publisher={IEEE Computer Society},
}

@book{bangerth2003adaptive,
    title={Adaptive finite element methods for differential equations},
    author={Bangerth, Wolfgang and Rannacher, Rolf},
    year={2003},
    publisher={Springer Science \& Business Media}
}

@article{bonito2024adaptive,
    title={Adaptive finite element methods},
    author={Bonito, Andrea and Canuto, Claudio and Nochetto, Ricardo H and Veeser, Andreas},
    journal={Acta Numerica},
    volume={33},
    pages={163--485},
    year={2024},
    publisher={Cambridge University Press}
}

@article{carstensen2006convergence,
    title={Convergence analysis of an adaptive nonconforming finite element method},
    author={Carstensen, Carsten and Hoppe, Ronald HW},
    journal={Numerische Mathematik},
    volume={103},
    number={2},
    pages={251--266},
    year={2006},
    publisher={Springer}
}

@article{Blayo1999ocean,
    author = "Eric Blayo and Laurent Debreu",
    title = "Adaptive Mesh Refinement for Finite-Difference Ocean Models: First Experiments",
    journal = "Journal of Physical Oceanography",
    year = "1999",
    publisher = "American Meteorological Society",
    volume = "29",
    number = "6",
    doi = "10.1175/1520-0485(1999)029<1239:AMRFFD>2.0.CO;2",
    pages= "1239--1250",
}

@article{essadki2016adaptive,
    title={Adaptive mesh refinement and high order geometrical moment method for the simulation of polydisperse evaporating sprays},
    author={Essadki, Mohamed and De Chaisemartin, St{\'e}phane and Massot, Marc and Laurent, Fr{\'e}d{\'e}rique and Larat, Adam and Jay, St{\'e}phane},
    journal={Oil \& Gas Science and Technology--Revue d’IFP Energies nouvelles},
    volume={71},
    number={5},
    pages={61},
    year={2016},
    publisher={EDP Sciences}
}

@article{lal2024accuracy,
    title={Accuracy verification of a 2D adaptive mesh refinement method by the benchmarks of lid-driven cavity flows with an arbitrary number of refinements},
    author={Lal, Rajnesh and Li, Zhenquan and Li, Miao},
    journal={Mathematics},
    volume={12},
    number={18},
    pages={2831},
    year={2024},
    publisher={MDPI}
}

@inproceedings{Losasso2004octree,
    author = {Losasso, Frank and Gibou, Fr\'{e}d\'{e}ric and Fedkiw, Ron},
    title = {Simulating water and smoke with an octree data structure},
    year = {2004},
    isbn = {9781450378239},
    publisher = {Association for Computing Machinery},
    doi = {10.1145/1186562.1015745},
    booktitle = {ACM SIGGRAPH 2004 Papers},
    pages = {457–462},
    numpages = {6},
    series = {SIGGRAPH '04}
}

@article{ando2013adaptive,
    author = {Ando, Ryoichi and Th\"{u}rey, Nils and Wojtan, Chris},
    title = {Highly adaptive liquid simulations on tetrahedral meshes},
    year = {2013},
    issue_date = {July 2013},
    publisher = {Association for Computing Machinery},
    volume = {32},
    number = {4},
    issn = {0730-0301},
    doi = {10.1145/2461912.2461982},
    journal = {ACM Trans. Graph.},
    month = jul,
    articleno = {103},
    numpages = {10},
}

@article{Setaluri2014adaptive,
    author = {Setaluri, Rajsekhar and Aanjaneya, Mridul and Bauer, Sean and Sifakis, Eftychios},
    title = {SPGrid: a sparse paged grid structure applied to adaptive smoke simulation},
    year = {2014},
    issue_date = {November 2014},
    publisher = {Association for Computing Machinery},
    volume = {33},
    number = {6},
    issn = {0730-0301},
    doi = {10.1145/2661229.2661269},
    journal = {ACM Trans. Graph.},
    month = nov,
    articleno = {205},
    numpages = {12},
}

@article{Raateland2022dcgrid,
    author = {Raateland, Wouter and H\"{a}drich, Torsten and Herrera, Jorge Alejandro Amador and Banuti, Daniel T. and Pa\l{}ubicki, Wojciech and Pirk, S\"{o}ren and Hildebrandt, Klaus and Michels, Dominik L.},
    title = {DCGrid: An Adaptive Grid Structure for Memory-Constrained Fluid Simulation on the GPU},
    year = {2022},
    issue_date = {May 2022},
    publisher = {Association for Computing Machinery},
    volume = {5},
    number = {1},
    doi = {10.1145/3522608},
    journal = {Proc. ACM Comput. Graph. Interact. Tech.},
    articleno = {3},
    numpages = {14},
}

@book{ferziger2019computational,
    title={Computational methods for fluid dynamics},
    author={Ferziger, Joel H and Peri{\'c}, Milovan and Street, Robert L},
    volume={3},
    year={2019},
    publisher={springer}
}

@book{schwarz2006hodge,
    title={Hodge Decomposition-A method for solving boundary value problems},
    author={Schwarz, G{\"u}nter},
    year={2006},
    publisher={Springer}
}

@book{verfurth2013posteriori,
    title={A posteriori error estimation techniques for finite element methods},
    author={Verf{\"u}rth, R{\"u}diger},
    year={2013},
    publisher={OUP Oxford}
}

@article{verfurth2010posteriori,
    title={A posteriori error analysis of space-time finite element discretizations of the time-dependent Stokes equations},
    author={Verf{\"u}rth, R{\"u}diger},
    journal={Calcolo},
    volume={47},
    number={3},
    pages={149--167},
    year={2010},
    publisher={Springer}
}

@article{doerfler_1996_convergent,
    author = {D{\"o}rfler, Willy},
    title = {A Convergent Adaptive Algorithm for Poisson’s Equation},
    journal = {SIAM Journal on Numerical Analysis},
    volume = {33},
    number = {3},
    pages = {1106-1124},
    year = {1996},
    doi = {10.1137/0733054},
}

@article{stevenson2007optimality,
    title={Optimality of a standard adaptive finite element method},
    author={Stevenson, Rob},
    journal={Foundations of Computational Mathematics},
    volume={7},
    number={2},
    pages={245--269},
    year={2007},
    publisher={Springer}
}

@book{mitchell1988unified,
    title={Unified multilevel adaptive finite element methods for elliptic problems},
    author={Mitchell, William F},
    year={1988},
    publisher={University of Illinois at Urbana-Champaign}
}

@article{mitchell1989comparison,
    author = {Mitchell, William F.},
    title = {A comparison of adaptive refinement techniques for elliptic problems},
    year = {1989},
    publisher = {Association for Computing Machinery},
    volume = {15},
    number = {4},
    issn = {0098-3500},
    doi = {10.1145/76909.76912},
    journal = {ACM Trans. Math. Softw.},
    pages = {326–347},
    numpages = {22}
}

@article{chen2010coarsening,
    title={A coarsening algorithm on adaptive grids by newest vertex bisection and its applications},
    author={Chen, Long and Zhang, Chensong},
    journal={Journal of Computational Mathematics},
    pages={767--789},
    year={2010},
    publisher={JSTOR}
}

@article{Biedl2001efficient,
    title = {Efficient Algorithms for Petersen's Matching Theorem},
    author = {Therese C. Biedl and Prosenjit Bose and Erik D. Demaine and Anna Lubiw},
    journal = {Journal of Algorithms},
    volume = {38},
    number = {1},
    pages = {110-134},
    year = {2001},
    issn = {0196-6774},
    doi = {https://doi.org/10.1006/jagm.2000.1132},
}

@article{KOSSACZKY1994recursive,
    title = {A recursive approach to local mesh refinement in two and three dimensions},
    author = {Igor Kossaczký},
    journal = {Journal of Computational and Applied Mathematics},
    volume = {55},
    number = {3},
    pages = {275-288},
    year = {1994},
    issn = {0377-0427},
    doi = {https://doi.org/10.1016/0377-0427(94)90034-5},
}

@article{Sharp2019intrinsic,
    author = {Sharp, Nicholas and Soliman, Yousuf and Crane, Keenan},
    title = {Navigating intrinsic triangulations},
    year = {2019},
    issue_date = {August 2019},
    publisher = {Association for Computing Machinery},
    volume = {38},
    number = {4},
    issn = {0730-0301},
    doi = {10.1145/3306346.3322979},
    journal = {ACM Trans. Graph.},
    month = jul,
    articleno = {55},
    numpages = {16},
}

@article{Gillespie2021intrinsic,
    author = {Gillespie, Mark and Sharp, Nicholas and Crane, Keenan},
    title = {Integer coordinates for intrinsic geometry processing},
    year = {2021},
    issue_date = {December 2021},
    publisher = {Association for Computing Machinery},
    volume = {40},
    number = {6},
    issn = {0730-0301},
    doi = {10.1145/3478513.3480522},
    journal = {ACM Trans. Graph.},
}

@article{DormandPrince1980,
    title = {A family of embedded Runge-Kutta formulae},
    author = {J.R. Dormand and P.J. Prince},
    journal = {Journal of Computational and Applied Mathematics},
    volume = {6},
    number = {1},
    pages = {19-26},
    year = {1980},
    issn = {0377-0427},
    doi = {https://doi.org/10.1016/0771-050X(80)90013-3},
}

@book{hairer1993solving,
    title={Solving ordinary differential equations I: Nonstiff problems},
    author={Hairer, Ernst and Wanner, Gerhard and N{\o}rsett, Syvert P},
    year={1993},
    publisher={Springer}
}

@article{stevenson2008completion,
    title={The completion of locally refined simplicial partitions created by bisection},
    author={Stevenson, Rob},
    journal={Mathematics of computation},
    volume={77},
    number={261},
    pages={227--241},
    year={2008}
}

@inproceedings{erickson2005greedy,
    title={Greedy optimal homotopy and homology generators},
    author={Erickson, Jeff and Whittlesey, Kim},
    booktitle={SODA},
    volume={5},
    pages={1038--1046},
    year={2005}
}

@article{dlotko2012fast,
    title = {A fast algorithm to compute cohomology group generators of orientable 2-manifolds},
    author = {Paweł Dłotko},
    journal = {Pattern Recognition Letters},
    volume = {33},
    number = {11},
    pages = {1468-1476},
    year = {2012},
    note = {Computational Topology in Image Context},
    issn = {0167-8655},
    doi = {https://doi.org/10.1016/j.patrec.2011.10.005},
}

@article{eppstein2002generators,
    author       = {David Eppstein},
    title        = {Dynamic Generators of Topologically Embedded Graphs},
    journal      = {CoRR},
    volume       = {cs.DS/0207082},
    year         = {2002},
    url          = {https://arxiv.org/abs/cs/0207082},
    bibsource    = {dblp computer science bibliography, https://dblp.org}
}

@article{dey1998homology,
    author = {Dey, Tamal K. and Guha, Sumanta},
    title = {Computing homology groups of simplicial complexes in R3},
    year = {1998},
    publisher = {Association for Computing Machinery},
    volume = {45},
    number = {2},
    issn = {0004-5411},
    doi = {10.1145/274787.274810},
    journal = {J. ACM},
    month = mar,
    pages = {266–287},
    numpages = {22},
}

@book{bossavit1998computational,
    title={Computational electromagnetism: variational formulations, complementarity, edge elements},
    author={Bossavit, Alain},
    year={1998},
    publisher={Academic Press}
}

@article{Hiptmair_2002,
    title={Finite elements in computational electromagnetism},
    author={Hiptmair, R.},
    year={2002},
    journal={Acta Numerica},
    pages={237–339},
    volume={11},
    DOI={10.1017/S0962492902000041},
}

@article{edelsbrunner2020tri,
    title={Tri-partitions and bases of an ordered complex},
    author={Edelsbrunner, Herbert and {\"O}lsb{\"o}ck, Katharina},
    journal={Discrete \& Computational Geometry},
    volume={64},
    number={3},
    pages={759--775},
    year={2020},
    publisher={Springer}
}

@incollection{crane2013digital,
    title={Digital geometry processing with discrete exterior calculus},
    author={Crane, Keenan and De Goes, Fernando and Desbrun, Mathieu and Schr{\"o}der, Peter},
    booktitle={ACM SIGGRAPH 2013 Courses},
    pages={1--126},
    year={2013}
}

@article{Dupont_BFECC,
    title = {Back and forth error compensation and correction methods for removing errors induced by uneven gradients of the level set function},
    author = {Todd F. Dupont and Yingjie Liu},
    journal = {Journal of Computational Physics},
    volume = {190},
    number = {1},
    pages = {311-324},
    year = {2003},
    issn = {0021-9991},
    doi = {https://doi.org/10.1016/S0021-9991(03)00276-6},
}

@inproceedings{kim2005flowfixer,
    title={FlowFixer: Using BFECC for Fluid Simulation.},
    author={Kim, ByungMoon and Liu, Yingjie and Llamas, Ignacio and Rossignac, Jarek},
    booktitle={NPH},
    pages={51--56},
    year={2005}
}

@article{geometrycentral,
    title={GeometryCentral: A modern C++ library of data structures and algorithms for geometry processing},
    author={Nicholas Sharp and Keenan Crane and others},
    howpublished="\url{https://geometry-central.net/}",
    year={2019}
}

@MISC{eigenweb,
    author = {Ga\"{e}l Guennebaud and Beno\^{i}t Jacob and others},
    title = {Eigen},
    howpublished = {https://libeigen.gitlab.io},
    year = {2010}
}

% biblatex with biber
% \printbibliography

%-------------------------------------------------------------------------
%Color tables are no longer required for purely electronic publications.
%\newpage

\end{document}